\documentclass[10pt]{article}
\usepackage{amsmath}
\usepackage{graphicx}
\usepackage{color}
\usepackage{orcidlink}
\usepackage{hyperref}
\hypersetup{colorlinks=true, linkcolor=blue, citecolor=blue, urlcolor=blue}
\usepackage{caption}
\usepackage{subcaption}
\usepackage{setspace}
\usepackage{multirow}
\usepackage{multicol}
\usepackage{amssymb}
\RequirePackage[numbers,sort&compress]{natbib}
\begin{document}
\baselineskip=18pt

\begin{center}
\LARGE{\bf Dymnikova black hole immersed in perfect fluid dark matter and a cloud of strings: Hawking Temperature, Dynamics and QPOs Analysis}
\end{center}

\vspace{0.2cm}

\begin{center}
{\bf Faizuddin Ahmed\orcidlink{0000-0003-2196-9622}}\footnote{\bf faizuddinahmed15@gmail.com }\\
{\it Department of Physics, The Assam Royal Global University, Guwahati, 781035, Assam, India}\\

{\bf Sardor Murodov\orcidlink{0000-0003-2360-4475}}\footnote{\bf s.murodov@newuu.uz (Corresp. author)}\\{\it New Uzbekistan University, Movarounnahr Str. 1, Tashkent 100000, Uzbekistan}\\{\it Institute of Fundamental and Applied Research, National Research University TIIAME, Kori Niyoziy 39, Tashkent 100000, Uzbekistan}\\

{\bf Bekzod Rahmatov\orcidlink{0009-0001-0394-650X}}\footnote{\bf be.rahmatov@newuu.uz}\\{\it University of Tashkent for Applied Sciences, Str. Gavhar 1, Tashkent 100149, Uzbekistan}\\{\it Tashkent State Technical University, Tashkent 100095, Uzbekistan}\\

{\bf Abdelmalek Bouzenada\orcidlink{0000-0002-3363-980X}}\footnote{\bf abdelmalekbouzenada@gmail.com}\\ 
{\it Laboratory of Theoretical and Applied Physics, Echahid Cheikh Larbi Tebessi University 12001, Algeria}\\
\&\\
{\it Research Center of Astrophysics and Cosmology, Khazar University, Baku, AZ1096, 41 Mehseti Street, Azerbaijan}\\

\end{center}

\date{\today}

\vspace{0.5cm}

\begin{abstract}
The Dymnikova black hole is a regular spacetime solution that smoothly interpolates between a de Sitter–like core near the origin and a Schwarzschild-type geometry at asymptotically large distances. In this work, we examine a generalized Dymnikova black hole surrounded by perfect fluid dark matter and immersed in a cloud of strings, and investigate how these additional matter fields modify the physical properties of the spacetime. We carry out a detailed analysis of the thermodynamic behavior, optical characteristics, and dynamical aspects of test particle motion, with particular emphasis on quasiperiodic oscillations (QPOs). Our results show that the presence of perfect fluid dark matter and the string cloud significantly affects the Hawking temperature and the specific heat of the black hole, leading to rich thermodynamic behavior and parameter-dependent phase transitions. We further demonstrate that the photon sphere and the corresponding black hole shadow are highly sensitive to variations in the model parameters, resulting in observable deviations from the standard Dymnikova and Schwarzschild cases. In addition, we derive key thermodynamic quantities, including the Hawking temperature and heat capacity, and analyze how the underlying geometric parameters influence the stability and phase structure of the black hole. Finally, we study quasiperiodic oscillations arising from particle motion in the vicinity of the black hole and show how the characteristic frequencies encode signatures of perfect fluid dark matter and the string cloud.
\end{abstract}

\pagebreak

\tableofcontents

\section{Introduction}

The formulation of general relativity (GR) illustrated a fundamental shift in gravitational physics, using a geometric description of spacetime in which gravitation arises from curvature rather than force interactions \cite{BHM1}. Shortly thereafter, the exact solution describing a point-like mass demonstrated the presence of an extreme spacetime structure characterized by a nontrivial causal boundary that prevents signal escape \cite{BHM2}. Such solutions, subsequently embedded within the consistent theoretical framework of relativistic gravitation, are presently identified as BHs, which are defined by event horizons and, in generic situations, curvature singularities at the core \cite{BHM3}. With continued theoretical and observational progress, BHs transitioned from mathematical predictions to astrophysical entities, a development strongly confirmed by the direct observation of gravitational wave signals emitted during binary BH coalescences \cite{BHM4, BHM10}. Also, this empirical advance was further strengthened by horizon-scale observations of supermassive BHs, yielding direct measurements of shadow morphology consistent with relativistic predictions \cite{BHM11,BHM16}. A major revision of the classical paradigm emerged when quantum considerations revealed that BHs emit thermal radiation, thereby possessing well-defined temperature and entropy \cite{BHM17}. 

Regular BHs (RBHs) are characterized as BH solutions that admit event horizons while excluding essential spacetime singularities. Initial constructions of these configurations were presented by Dymnikova \cite{RBH1}, after which consistent field-theoretical descriptions based on nonlinear electrodynamics were formulated by Ay\'on-Beato and Garc\'ia \cite{RBH2} and subsequently generalized by Bronnikov \cite{RBH3}. Fundamental aspects of singularity formation and the behavior of curvature in general relativity were analyzed systematically in classical studies \cite{RBH4}. The requirement that curvature invariants remain globally bounded was later articulated through Markov’s limiting curvature conjecture \cite{RBH5}, which motivated extensive investigations in BH models \cite{RBH6, RBH7, RBH8}. Nevertheless, the Taub-NUT spacetime constitutes a notable counterexample, since all curvature invariants are finite, whereas geodesic incompleteness is still present \cite{RBH9, RBH10}. The notion of geodesic completeness itself is linked to the global causal structure of spacetime, as discussed in standard treatments \cite{RBH11}, although it has been demonstrated that space-times satisfying geodesic completeness may nevertheless possess divergent curvature invariants \cite{RBH13, RBH14}, consistent with earlier conceptual examinations of singularities \cite{RBH15, RBH16}. The physical motivation underlying RBHs originates from the pioneering proposals of Sakharov and Gliner, who suggested substituting the classical vacuum by a vacuum-like state \cite{RBH17, RBH18}, an idea later implemented within cosmological and gravitational frameworks \cite{RBH19, RBH20, RBH21}. Detailed assessments of these developments are available in several review articles \cite{RBH22, RBH23}. In this case, the earliest explicit RBH metric was constructed by Bardeen \cite{RBH24}, with its interpretation in terms of nonlinear electrodynamics provided subsequently in \cite{RBH25}. Also, this approach has been systematically generalized to obtain a wide family of RBHs \cite{RBH26, RBH28}, thereby establishing nonlinear electrodynamics as a primary mechanism for realizing regular BH geometries. 

Contemporary observational cosmology establishes that the current universe is predominantly governed by non-baryonic sectors, while baryonic matter contributes only a minor fraction relative to dark matter and dark energy \cite{DMS1}. Such empirical results strongly motivate investigations of BH space-times immersed in environments modified by dark components, as realistic astrophysical systems inevitably interact with cosmological background fields. Within this context, BHs embedded in quintessence-like dark energy distributions have been studied extensively. The first formulation of a Schwarzschild BH coupled to a quintessential field was introduced in \cite{DMS2}, followed by extensions to rotating configurations in \cite{DMS3}. In this case, later analyses explored multiple physical characteristics, such as quasinormal spectra, thermodynamic behavior, and phase structures of regular BHs under quintessence effects \cite{DMS4, DMS5}. In parallel, perfect fluid dark matter (PFDM) emerged as an effective phenomenological model that successfully accounts for the flat rotation curves observed in spiral galaxies \cite{DMS2, DMS11}. This model has been extensively examined for both static and rotating BH geometries, particularly with respect to optical signatures, shadow formation, and test-particle motion \cite{Shermatov2025, Chaudhary2025, Jumaniyozov2025}. Also, inspired by these results, and building upon previous formulations of regular BHs within dark energy environments, we generalize the Schwarzschild BH surrounded by PFDM to a spherically symmetric Bardeen BH. Moreover, employing the Newman-Janis algorithm, we construct the associated rotating Bardeen BH solution embedded in PFDM, yielding a more astrophysical relevant framework for studying BHs affected by dark matter distributions. 

The cloud of strings model has been extensively employed as an alternative source of matter in gravitational frameworks, wherein the spacetime is assumed to be filled with one-dimensional strings that can span cosmological distances. In this case, the first exact solution of Einstein’s field equations, including a quintessence field in four dimensions, was derived by Kiselev \cite{CS1}, and later extensions to higher-dimensional ($\mathcal{D})$ geometries were obtained by Chen et al. \cite{CS2}. Within the thin-shell wormhole approach, Banerjee et al. \cite{CS3} analyzed the dynamical stability of d-dimensional constructions in the presence of a quintessence parameter, showing how this parameter affects the throat stability conditions. Apart from quintessence, cosmic strings provide an alternative mechanism to model exotic matter, since they correspond to one-dimensional topological defects that may have been produced during early-universe phase transitions and can influence the late-time accelerated expansion \cite{CS4}. Motivated by string theory, where fundamental particles and interactions are described as vibrational excitations of microscopic supersymmetric strings, several works have focused on the gravitational implications of string clouds. Letelier \cite{PSL1979} initiated this research by obtaining general spherically symmetric solutions for clouds of strings and by discussing the relation between string state counting and BH entropy.

In this work, we aim to study thermodynamic and dynamical properties of a Dymnikova black hole immersed in both a perfect fluid dark matter (PFDM) and a cloud of strings (CS). The presence of these external matter sources modifies the spacetime geometry and consequently affects both the thermodynamic behavior and the motion of test particles in the black hole background. First, we analyze the thermodynamic properties of the system by deriving the Hawking temperature and the specific heat capacity. These quantities enable us to explore the thermal stability of the black hole and to examine how the PFDM parameter and the cloud of strings parameter influence its phase structure and stability conditions. Next, we turn to the study of photon dynamics in this modified spacetime. In particular, we determine the photon sphere radius and investigate the corresponding black hole shadow. The effects of the PFDM and CS parameters on the radius of unstable circular photon orbits and on the size and shape of the shadow are analyzed in detail, providing insight into potential observational signatures. We further examine the dynamics of massive test particles by studying timelike geodesics in the considered background. Special attention is devoted to circular orbits and their stability properties. The influence of the dark matter distribution and the string cloud on the effective potential and orbital structure is discussed. In addition, we investigate quasi-periodic oscillations (QPOs) by computing the fundamental frequencies associated with particle motion, namely the radial, azimuthal, and vertical epicyclic frequencies. We analyze how the PFDM and CS parameters modify these frequencies and discuss their possible astrophysical implications.

The structure of this paper is organized as follows. In Sec.~\ref{sec:2}, we present the background setup and describe the spacetime geometry of the Dymnikova black hole in the presence of PFDM and a cloud of strings. In Sec.~\ref{sec:3}, we derive the Hawking temperature and analyze the specific heat capacity, focusing on thermodynamic stability. In Secs.~\ref{sec:4} and \ref{sec:5}, we study the dynamics of massless and massive test particles, respectively. In Sec.~\ref{sec:6}, we analyze quasi-periodic oscillations by computing the radial, azimuthal, and vertical epicyclic frequencies. In Sec.~\ref{sec:7}, we compare our theoretical results with available observational data. Finally, in Sec.~\ref{sec:8}, we summarize our findings and present concluding remarks.

\section{Dymnikova black hole immersed in PFDM and CS}\label{sec:2}

Among the known regular black hole models, such as the Bardeen, Hayward, Frolov, and Simpson-Visser solutions, another important regular solution was proposed by Dymnikova, based on the gravitational analogue of the Schwinger effect \cite{ID}. This gravitational analogue represents a vacuum solution that, at large distances from the origin, behaves like the Schwarzschild solution, while near the core, it approaches the De Sitter solution. Recently, the Dymnikova black hole has been studied in various contexts, such as pure Lovelock gravity \cite{ME2025}, the generalized uncertainty principle \cite{DM2024}, theories with higher-curvature corrections \cite{RAK2024}, and wormhole solutions \cite{ME2023}. Notably, the quasinormal modes and thermodynamic properties of the Dymnikova black hole in higher dimensions have been investigated in \cite{MHM2024}. Recently, the properties of a Dymnikova black hole immersed in a quintessential field has investigated in \cite{MHM2026}. In this paper, we consider the Dymnikova black hole immersed in a perfect fluid dark matter and surrounded by a cloud of strings. We investigate how these external factors affect the thermodynamic and dynamical properties of the black hole, comparing the results to the corresponding properties of the Dymnikova regular black hole model.

The metric corresponding to the Dymnikova black hole is given by \cite{ID}
\begin{equation}
ds^2 = -h(r)dt^2 + \dfrac{dr^2}{h(r)} + r^2(d\theta ^2 + \sin ^2{\theta }\hspace{0.1cm}d\varphi ^2),\label{dymnikova}
\end{equation}
where 
\begin{equation}
     h(r) = 1- \dfrac{r_g}{r}\left(1 - e^{-r^{3}/r_{*}^3}\right)\label{dymnikova2}
\end{equation}
and $r_{*}^3 = r_g r_0^2$ with $r_g=2\,M$ denoting the Schwarzschild radius and $r_0$ the de Sitter radius.

Since we are mainly interested on dark matter as a type of perfect fluid, the energy-momentum of perfect fluid dark matter is given by \cite{MHL2012,HXZ2021}
\begin{equation}
T^{\text{DM}}_{\,\,\mu\nu} = \mathrm{diag}\left(-\mathcal{E}_{DM},\, P_{r\,DM},\, P_{\theta\,DM},\, P_{\phi\,DM}\right),
\label{tensor}
\end{equation}
where the components are
\begin{equation}
\mathcal{E}_{DM} = -P_{r\,DM}= -\frac{\lambda}{8\pi r^3},\qquad P_{\theta\,DM}= -\frac{\lambda}{16\pi r^3}=P_{\phi\,DM}.
\label{energy-density}
\end{equation}
Here $\lambda$ is a constant taking real values called PFDM parameter.

To include string-like objects, we consider Nambu-Goto action given by \cite{PSL1979}
\begin{equation}
    S_{\rm CS}=\int \sqrt{-\gamma}\,\mathcal{M}\,d\lambda^0\,d\lambda^1=\int \mathcal{M}\sqrt{-\frac{1}{2}\,\Sigma^{\mu \nu}\,\Sigma_{\mu\nu}}\,d\lambda^0\,d\lambda^1,\label{act1}
\end{equation}
where $\mathcal{M}$ is the dimensionless constant which characterizes the string, ($\lambda^0\,\lambda^1$) are the time
like and spacelike coordinate parameters, respectively \cite{JLS1960}. $\gamma$  is the determinant of the induced metric of the strings world sheet given by $\gamma=g^{\mu\nu}\frac{ \partial x^\mu}{\partial \lambda^a}\frac{ \partial x^\nu}{\partial \lambda^b}$.  $\Sigma_{\mu\nu}=\epsilon^{ab}\frac{ \partial x^\mu}{\partial \lambda^a}\frac{ \partial x^\nu}{\partial \lambda^b}$ is bivector related to string world sheet, where $\epsilon^{ab}$ is the second rank Levi-Civita tensor which takes the non-zero values as $\epsilon^{01} = -\epsilon^{10} = 1$.

\begin{equation}
   T_{\mu\nu}^{\rm CS}=2 \frac{\partial}{\partial g_{\mu \nu}}\mathcal{M}\sqrt{-\frac{1}{2}\Sigma^{\mu \nu}\,\Sigma_{\mu\nu}} =\frac{\rho^{\rm CS} \,\Sigma_{\alpha\nu}\, \,\Sigma_{\mu}^\alpha }{\sqrt{-\gamma}}, \label{act2}
 \end{equation}
where $\rho^{\rm CS}$ is the proper density of the string cloud. The energy-momentum tensor components are given by
\begin{equation}
    T^{t\,(\rm CS)}_{t}=\rho^{\rm CS}=\frac{\alpha}{r^2}=T^{r\,(\rm CS)}_{r},\quad T^{\theta\,(\rm CS)}_{\theta}=T^{\phi\,(\rm CS)}_{\phi}= 0,\label{act3}
\end{equation}
where $\alpha$ is a constant associated with string-like objects called string parameter. Numerous researchers have studies black hole solution in Einstein gravity as well as modified gravity with CS in the literature (see, \cite{FA1,FA2,FA3,FA4}).

Thereby, incorporating both the perfect fluid dark matter and a cloud of strings into the Dymnikov black hole which is a vacuum solution, the metric corresponding to the new black hole is described by the following line-element 
\begin{equation}
    ds^2 = -f(r)dt^2 + \dfrac{dr^2}{f(r)} + r^2(d\theta ^2 + \sin ^2{\theta }\hspace{0.1cm}d\varphi ^2),\label{metric}
\end{equation}
where the lapse function is given by
\begin{equation}
     f(r) = 1-\alpha - \dfrac{r_g}{r}\left(1 - e^{-r^{3}/r_{*}^3}\right)+\dfrac{\lambda}{r}\ln\!\dfrac{r}{|\lambda|}.\label{function}
\end{equation}

\begin{figure}
    \centering
    \includegraphics[width=0.45\linewidth]{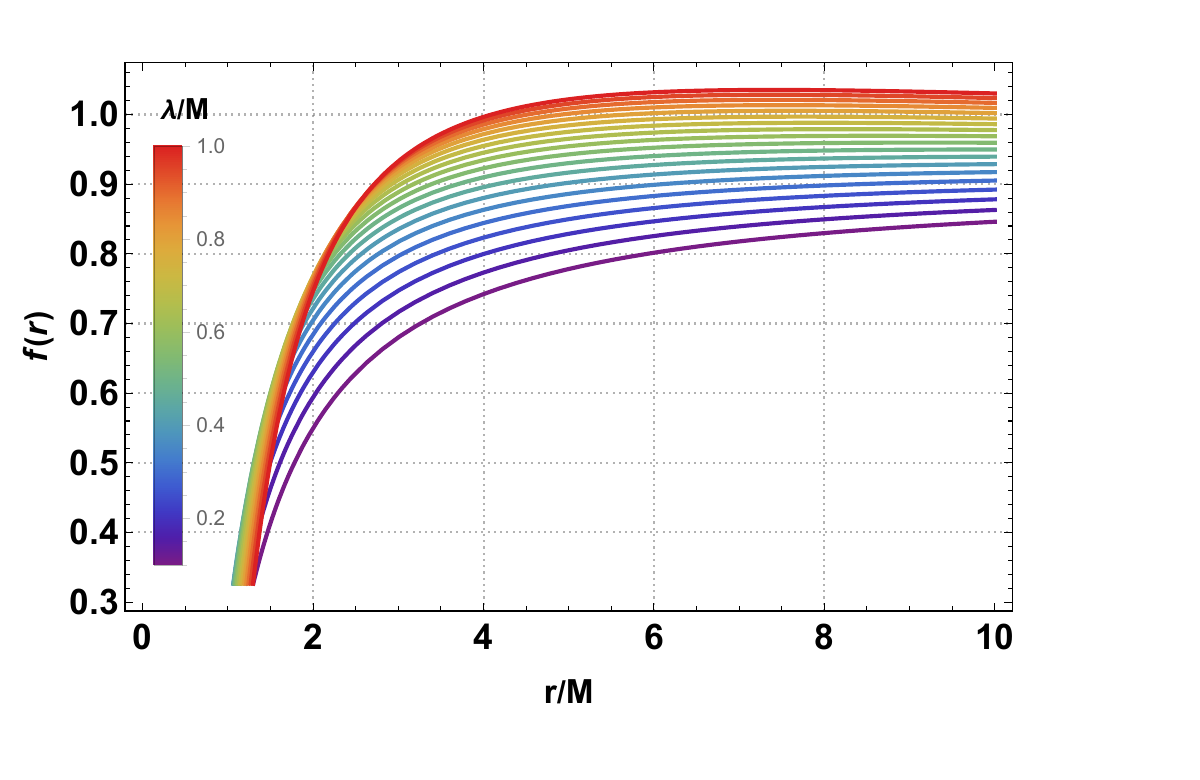}
    \includegraphics[width=0.45\linewidth]{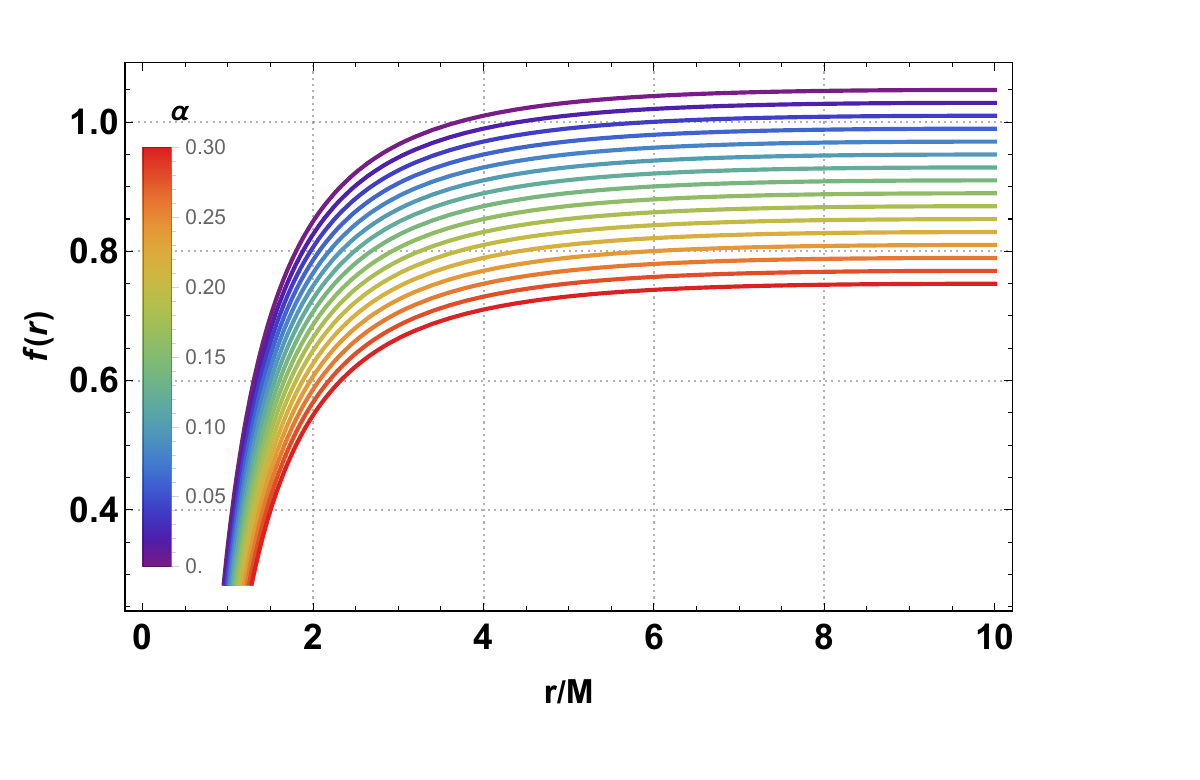}\\
    (i) $\alpha=0.1$ \hspace{8cm} (ii) $\lambda=0.5$
    \caption{The behavior of the metric function $f(r)$ as a function of dimensionless radial distance $r/M$ by varying, respectively the PFDM and CS parameters. Here $r_0/M=0.4,\,$}
    \label{fig:metric}
\end{figure}

    In Fig \ref{fig:metric}, the metric function $f(r)$ is presented as a function of the dimensionless radial coordinate $r/M$ for different values of the PFDM parameter $\lambda$ and the CS parameter $\alpha$, with $r_0/M = 0.4$. The curves demonstrate that $f(r)$ monotonically approaches unity as $r/M$ increases, confirming asymptotic flatness and the recovery of the classical general relativistic limit at large distances. The influence of the parameters is primarily concentrated in the small and intermediate radial regions, corresponding to the strong–field regime. An increase in the PFDM parameter $\lambda$ alters the curvature structure and modifies the radial behavior of the metric function, while a larger CS parameter $\alpha$ shifts $f(r)$ downward, indicating a deeper effective gravitational potential. Overall, both parameters introduce significant deviations from the Schwarzschild geometry in the strong gravitational region, whereas the asymptotic behavior remains consistent with general relativity.

Below, we discuss a few special cases.
\begin{itemize}
    \item In the asymptotic region $r \gg r_\ast$, the lapse function reduces to
    \begin{equation}
        f(r) = 1 - \alpha - \frac{r_g}{r}
        + \frac{\lambda}{r}\ln\!\frac{r}{|\lambda|}.
        \label{function1}
    \end{equation}
    In this limit, the spacetime described by Eq.~(\ref{metric}) corresponds to the
    Letelier black hole surrounded by perfect fluid dark matter \cite{AS2024}.

    \item In the near-region $r \ll r_\ast$, the metric function simplifies to
    \begin{equation}
        f(r) = 1 - \alpha - \Lambda r^2
        + \frac{\lambda}{r}\ln\!\frac{r}{|\lambda|},
        \qquad \Lambda = \frac{1}{r_0^{\,3}}.
        \label{function2}
    \end{equation}
    In this case, the space-time described by Eq.~(\ref{metric}) represents a de~Sitter core solution immersed in PFDM and CS.
\end{itemize}

\begin{figure}
    \centering
    \includegraphics[width=0.5\linewidth]{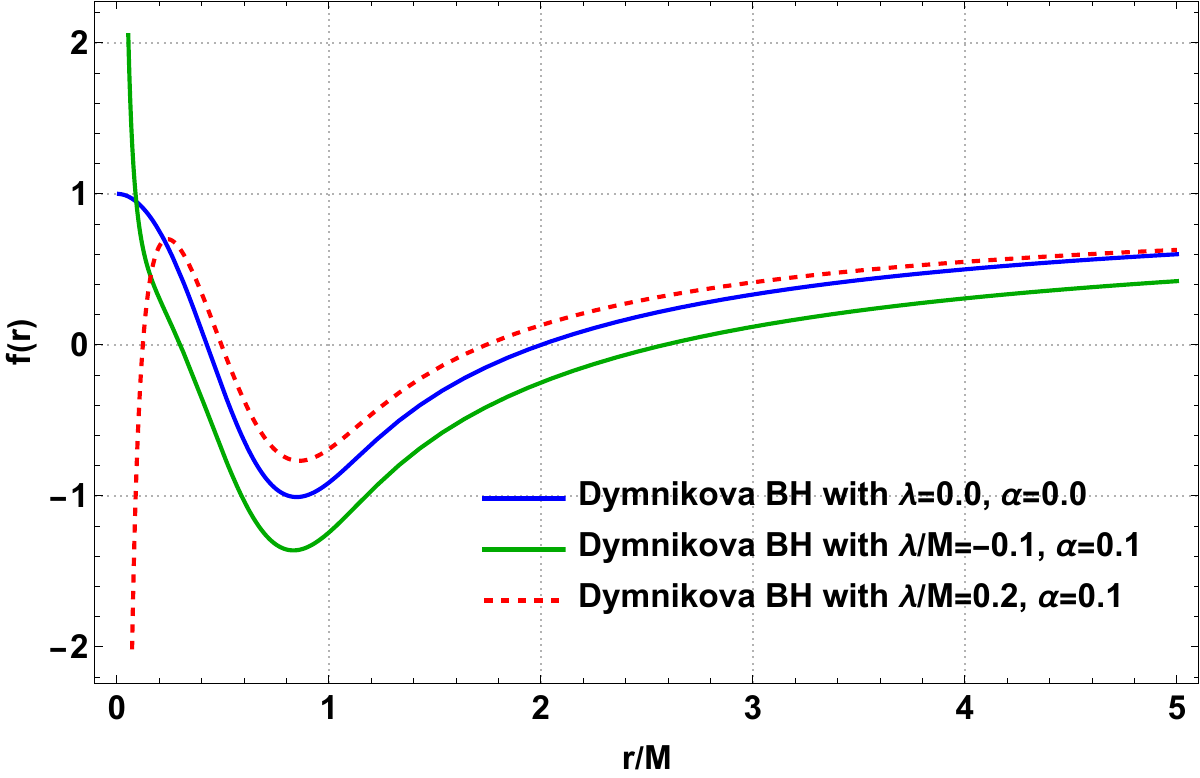}
    \caption{The metric function under various conditions. Here $r_0/M=0.4$.}
    \label{fig:function2}
\end{figure}

In Fig \ref{fig:function2}, the metric function $f(r)$ is presented for different values of the PFDM parameter $\lambda$ and the deformation parameter $\alpha$, with $r_0/M = 0.4$. While all curves asymptotically approach unity at large $r/M$, confirming asymptotic flatness, significant deviations arise in the strong–field region. The inclusion of PFDM and spacetime deformation modifies the depth and shape of the effective gravitational potential, particularly near the central region. These effects alter the near–horizon geometry, whereas the asymptotic behavior remains consistent with general relativity.

\section{Hawking Temperature and Specific Heat}\label{sec:3}

Black hole thermodynamics provides a fundamental link between gravity, quantum mechanics, and statistical physics. Following the pioneering work of Bekenstein \cite{Bekenstein1973}, black holes are assigned an entropy proportional to the area of their event horizon, known as the Bekenstein entropy. Hawking further demonstrated that black holes emit thermal radiation with a characteristic temperature, now referred to as the Hawking temperature \cite{Hawking1974,Hawking1975}, implying that they are genuine thermodynamic objects with well-defined energy, temperature, and entropy. This framework establishes the four laws of black hole mechanics, analogous to the laws of classical thermodynamics, and provides deep insights into the quantum properties of spacetime \cite{Wald1994,Carlip2014}. Understanding these thermodynamic quantities is essential for modeling accretion physics and the dynamics of matter near black holes, which also influences observable phenomena such as quasi-periodic oscillations. 

We begin this section by exploring the thermodynamics of the Dymnikova black hole surrounded by perfect fluid dark matter and a cloud of strings. To achieve this, we first analyze the black hole mass at the event horizon radius $r_h$ and is given by
\begin{equation}
M = \frac{r_h}{2} \left[ 1 - \alpha + \frac{\lambda}{r_h} \ln \frac{r_h}{|\lambda|} + \frac{r_0^2}{r_h^2}\, W\!\Bigg( - \frac{r_h^2}{r_0^2} e^{- \frac{r_h^2}{r_0^2} \left(1 - \alpha + \frac{\lambda}{r_h} \ln \frac{r_h}{|\lambda|}\right) } \Bigg) \right]^{-1},\label{mass}
\end{equation}
where \(W\) is the Lambert Function \cite{JL}. 

\begin{itemize}
    \item When $\lambda = 0$, corresponding to the absence of PFDM, the black hole mass reduces to
    \begin{equation}
    M = \frac{r_h}{2} 
    \left[ 
    1 - \alpha + \frac{r_0^2}{r_h^2}\, 
    W\!\Bigg( - \frac{r_h^2}{r_0^2} 
    e^{- \frac{r_h^2}{r_0^2} \left(1 - \alpha\right) } \Bigg) 
    \right]^{-1}, \label{mass1}
    \end{equation}

    \item When $\alpha=0$, corresponding to the absence of string-like objects, the black hole mass reduces to 
    \begin{equation}
M = \frac{r_h}{2} \left[ 1+ \frac{\lambda}{r_h} \ln \frac{r_h}{|\lambda|} + \frac{r_0^2}{r_h^2}\, W\!\Bigg( - \frac{r_h^2}{r_0^2} e^{- \frac{r_h^2}{r_0^2} \left(1+ \frac{\lambda}{r_h} \ln \frac{r_h}{|\lambda|}\right) } \Bigg) \right]^{-1},\label{mass3}
\end{equation}

    \item When $\alpha = 0 = \lambda$, corresponding to the absence of both string clouds and PFDM, the black hole mass further simplifies to
    \begin{equation}
    M = \frac{r_h}{2} 
    \left[ 
    1 + \frac{r_0^2}{r_h^2}\, 
    W\!\Bigg( - \frac{r_h^2}{r_0^2} 
    e^{- \frac{r_h^2}{r_0^2}} \Bigg) 
    \right]^{-1}. \label{mass2}
    \end{equation}
\end{itemize}
The above expression (\ref{mass2}) is similar to the well-known Dymnikova black hole mass \cite{MHM}.

As the lapse function at radial infinity is asymptotically bounded due to the presence of string cloud, that is, $\lim_{r \to \infty} f(r) =1-\alpha$. Therefore, the Hawking temperature for a static distant observer is given by \cite{Hawking1974,Hawking1975,Bekenstein1973}
\begin{equation}
    T = \dfrac{\kappa }{2\pi } = \dfrac{\mathcal{N}}{4\pi }\left. \dfrac{df(r)}{dr}\right| _{r = r_h},\label{ff1} 
\end{equation}
where $\mathcal{N}$ is determined as,
\begin{equation}
    \mathcal{N}=1/\sqrt{\lim_{r \to \infty} f(r)}=(1-\alpha)^{-1/2}.
\end{equation}

Substituting the metric function and after simplification results
\begin{equation}
T=\frac{(1-\alpha )^{-\frac{1}{2}}}{4 \pi  r_0} \left(\frac{2M \left(1-e^{-\frac{{r_h}^3}{2 M{r_0}^2}}\right)}{{r_h}^2}-\frac{\lambda  \ln \left(\frac{{r_h}}{| \lambda | }\right)}{{r_h}^2}-\frac{3 {r_h} e^{-\frac{{r_h}^3}{2M {r_0}^2}}}{{r_0}^2}+\frac{\lambda }{{r_h}^2}\right) .\label{ff2}
\end{equation}

\begin{itemize}
    \item When $\lambda = 0$, corresponding to the absence of PFDM, the temperature reduces to
    \begin{equation}
    T=\frac{(1-\alpha )^{-\frac{1}{2}}}{4 \pi  r_0} \left(\frac{2M }{{r_h}^2}\left(1-e^{-\frac{{r_h}^3}{2 M{r_0}^2}}\right)-\frac{3 {r_h} e^{-\frac{{r_h}^3}{2M {r_0}^2}}}{{r_0}^2}\right)\ .
        \label{ff3}
    \end{equation}

    \item When $\alpha=0$, corresponding to the absence of string-like objects, the Hawking temperature simplifies as,
    \begin{equation}
T=\frac{1}{4 \pi  r_0} \left(\frac{2M \left(1-e^{-\frac{{r_h}^3}{2 M{r_0}^2}}\right)}{{r_h}^2}-\frac{\lambda  \ln \left(\frac{{r_h}}{| \lambda | }\right)}{{r_h}^2}-\frac{3 {r_h} e^{-\frac{{r_h}^3}{2M {r_0}^2}}}{{r_0}^2}+\frac{\lambda }{{r_h}^2}\right) .\label{ff2a}
\end{equation}

    \item When $\alpha = 0 = \lambda$, corresponding to the absence of both string clouds and PFDM, the black hole mass further simplifies to
    \begin{equation}
   T=\frac{1}{4 \pi  r_0} \left(\frac{2M }{{r_h}^2}\left(1-e^{-\frac{{r_h}^3}{2 M{r_0}^2}}\right)-\frac{3 {r_h} e^{-\frac{{r_h}^3}{2M {r_0}^2}}}{{r_0}^2}\right)\ .
   \label{ff4}
\end{equation}
\end{itemize}
The above expression (\ref{ff4}) is similar to the well-known Hawking temperature of Dymnikova black hole \cite{ID}.
\begin{figure}[ht!]
\centering
    \includegraphics[width=0.49\linewidth]{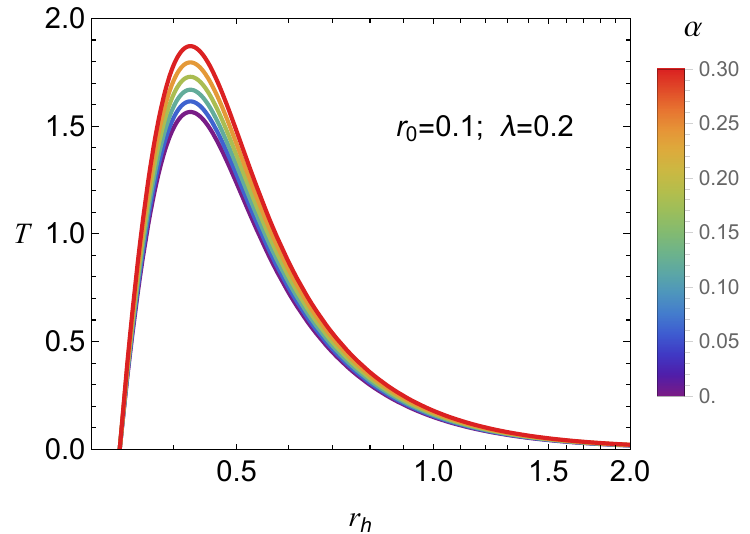}
    \includegraphics[width=0.49\linewidth]{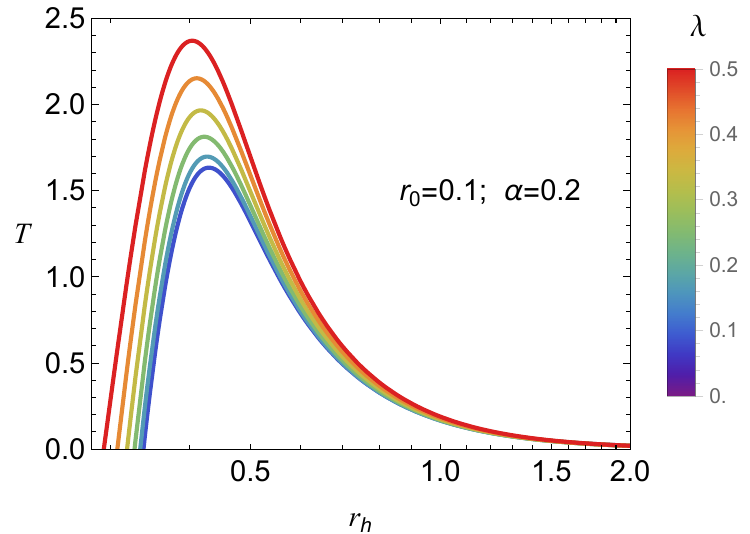}
    \includegraphics[width=0.49\linewidth]{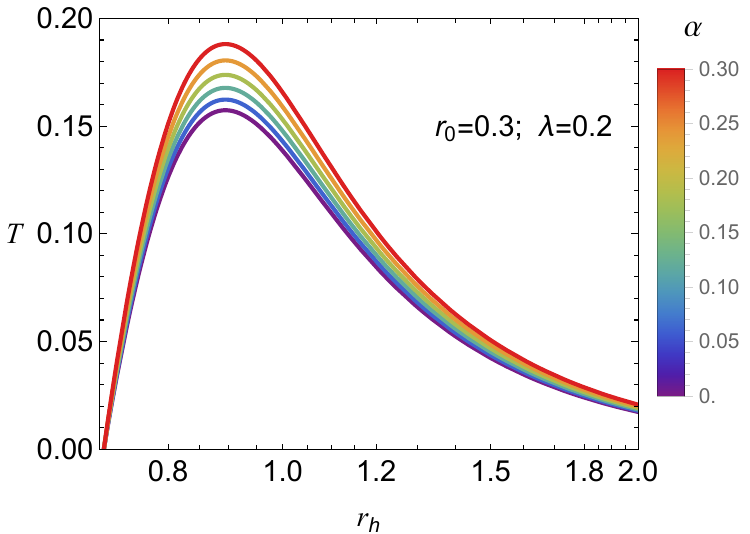}
    \includegraphics[width=0.49\linewidth]{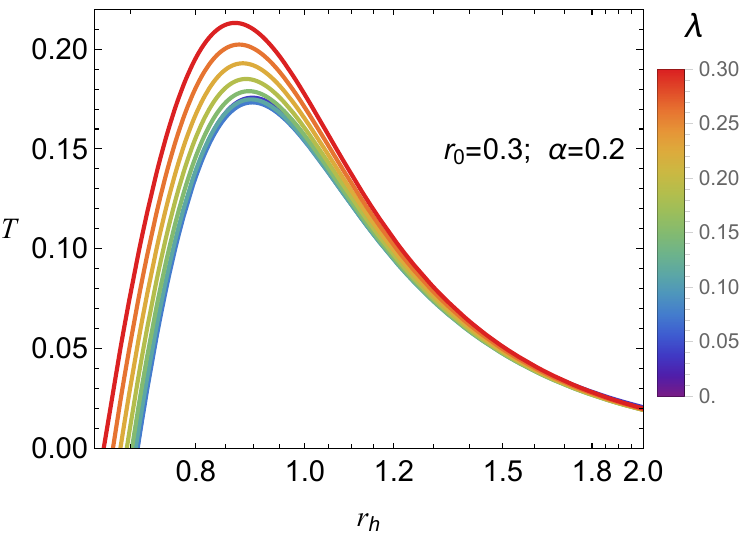}
   \caption{Hawking temperature $T$ as a function of the event horizon radius $r_h$ for different values of the model parameters. The upper row corresponds to $r_0 = 0.1$, while the lower row represents $r_0 = 0.3$. In the left panels, the parameter $\alpha$ is varied with fixed $\lambda$ in the right panels, $\lambda$ is varied with fixed $\alpha$. Different colors denote different parameter values.} \label{th}
\end{figure}
Figure~\ref{th} illustrates the dependence of the Hawking temperature $T$ on the event horizon radius $r_h$ for different values of the model parameters. In all cases, a non--monotonic behavior is observed: for small horizon radii the temperature increases rapidly, reaches a maximum, and then gradually decreases as $r_h$ grows. The presence of a single peak indicates the existence of a characteristic thermal scale, at which the black hole attains its highest radiation intensity.
An increase in the deformation parameter $\alpha$ shifts the temperature curves upward and enhances the maximal value of $T$. This behavior suggests that $\alpha$ modifies the spacetime geometry in such a way that the surface gravity near the horizon becomes stronger, leading to a higher thermal emission rate.
In contrast, increasing the PFDM parameter $\lambda$ suppresses the overall temperature and smooths the peak structure. This indicates that the surrounding dark matter distribution effectively reduces the surface gravity, thereby lowering the Hawking radiation intensity and moderating the thermal response of the system.
The scale parameter $r_0$ controls the global magnitude of the temperature. Larger values of $r_0$ lead to lower overall temperatures and shift the maximum toward larger horizon radii. This reflects the influence of the regular core structure, which softens the near--horizon geometry and reduces the rate at which the surface gravity varies.
For sufficiently large $r_h$, all curves tend to converge, showing that the influence of the deformation and dark matter parameters becomes less significant at large scales, where the thermodynamic behavior approaches the classical limit.

The determination of potential phase transitions in the black hole hinges on the criterion for a change in specific heat capacity sign. A positive specific heat capacity ($C>0$) is a signature of local stability against thermal fluctuations, whereas a negative specific heat capacity ($C<0$) indicates local instability. The expression for specific heat capacity  is as follows \cite{Wald1994,Carlip2014}:
\begin{equation}
C = \frac{dM}{dT}=\frac{\displaystyle \frac{4\pi M r_0 \sqrt{1-\alpha}}{r_h}\left[1 -\frac{\left\{- \dfrac{\lambda}{r_h}\!\left(1-\ln\!\frac{r_h}{|\lambda|}\right)- \dfrac{2 r_0^2}{r_h^2}W\!\left(- \dfrac{r_h^2}{r_0^2}e^{- \frac{r_h^2}{r_0^2} \left(1 - \alpha + \frac{\lambda}{r_h}\ln\frac{r_h}{|\lambda|}\right) }\right)+ \dfrac{r_0^2}{r_h}\dfrac{\partial W}{\partial r_h}\right\}}{\left\{1 -\alpha + \frac{\lambda}{r_h}\ln\frac{r_h}{|\lambda|} + \dfrac{r_0^2}{r_h^2}W\!\left(- \dfrac{r_h^2}{r_0^2}e^{- \frac{r_h^2}{r_0^2}\left(1 - \alpha + \frac{\lambda}{r_h}\ln\frac{r_h}{|\lambda|}\right) }\right)\right\}}\right]}{\displaystyle\left\{- \dfrac{r_0}{r_h^2}- \dfrac{3}{r_0}\left(1 - \dfrac{2 r_h}{r_g}\right)+ \dfrac{\lambda r_0}{r_h^3}\left(2 - \ln\!\frac{r_h}{|\lambda|}
\right)\right\}}.
\label{ff5}
\end{equation}

\begin{itemize}
\item When $\lambda = 0$, corresponding to the absence of PFDM, the specific heat reduces to
\begin{equation}
    C= \frac{\displaystyle \frac{4\pi M r_0 \sqrt{1-\alpha}}{r_h}
\left\{1 -\frac{\left[- \dfrac{2 r_0^2}{r_h^2}W\!\left(- \dfrac{r_h^2}{r_0^2}e^{- \frac{r_h^2}{r_0^2} (1-\alpha)}\right)
+ \dfrac{r_0^2}{r_h}\dfrac{\partial W}{\partial r_h}\right]}{\left[1 - \alpha+ \dfrac{r_0^2}{r_h^2}W\!\left(- \dfrac{r_h^2}{r_0^2}
e^{- \frac{r_h^2}{r_0^2} \left(1 - \alpha\right) }\right)\right]}\right\}}{\displaystyle \left\{- \dfrac{r_0}{r_h^2}- \dfrac{3}{r_0}
\left(1 - \dfrac{2 r_h}{r_g}\right)\right\}}.
\label{ff6}
\end{equation}

\item When $\alpha = 0$, corresponding to the absence of string-like objects, the specific heat further simplifies to
\begin{equation}
C =\frac{\displaystyle \frac{4\pi M r_0}{r_h}\left[1 -\frac{\left\{- \dfrac{\lambda}{r_h}\!\left(1-\ln\!\frac{r_h}{|\lambda|}\right)- \dfrac{2 r_0^2}{r_h^2}W\!\left(- \dfrac{r_h^2}{r_0^2}e^{- \frac{r_h^2}{r_0^2} \left(1+ \frac{\lambda}{r_h}\ln\frac{r_h}{|\lambda|}\right) }\right)+ \dfrac{r_0^2}{r_h}\dfrac{\partial W}{\partial r_h}\right\}}{\left\{1 + \frac{\lambda}{r_h}\ln\frac{r_h}{|\lambda|} + \dfrac{r_0^2}{r_h^2}W\!\left(- \dfrac{r_h^2}{r_0^2}e^{- \frac{r_h^2}{r_0^2}\left(1+ \frac{\lambda}{r_h}\ln\frac{r_h}{|\lambda|}\right) }\right)\right\}}\right]}{\displaystyle\left\{- \dfrac{r_0}{r_h^2}- \dfrac{3}{r_0}\left(1 - \dfrac{2 r_h}{r_g}\right)+ \dfrac{\lambda r_0}{r_h^3}\left(2 - \ln\!\frac{r_h}{|\lambda|}
\right)\right\}}.
\label{ff5a}
\end{equation}

\item When $\alpha = 0 = \lambda$, corresponding to the absence of both string clouds and PFDM, the specific heat further simplifies to
\begin{equation}
   C= \frac{\displaystyle \frac{4\pi M r_0}{r_h}\left\{1 -\frac{\left[- \dfrac{2 r_0^2}{r_h^2}W\!\left(- \dfrac{r_h^2}{r_0^2}e^{- \frac{r_h^2}{r_0^2}}\right)+ \dfrac{r_0^2}{r_h}\dfrac{\partial W}{\partial r_h}\right]}{\left[1+ \dfrac{r_0^2}{r_h^2}W\!\left(- \dfrac{r_h^2}{r_0^2}e^{- \frac{r_h^2}{r_0^2}}\right)\right]}\right\}}{\displaystyle\left\{- \dfrac{r_0}{r_h^2}- \dfrac{3}{r_0}\left(1 - \dfrac{2 r_h}{r_g}\right)\right\}}.\label{ff7}
\end{equation}
\end{itemize}

\begin{figure}
    \centering
    \includegraphics[width=0.45\linewidth]{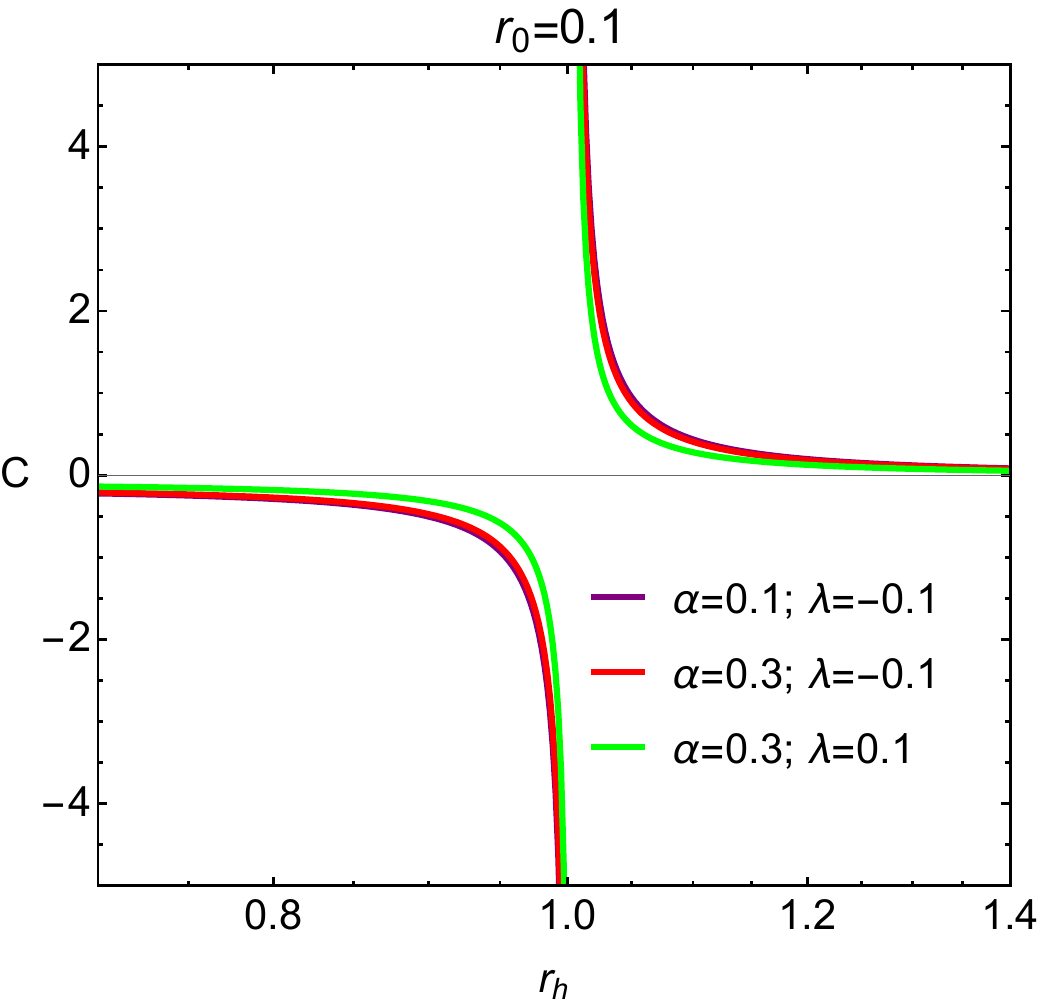}
    \includegraphics[width=0.45\linewidth]{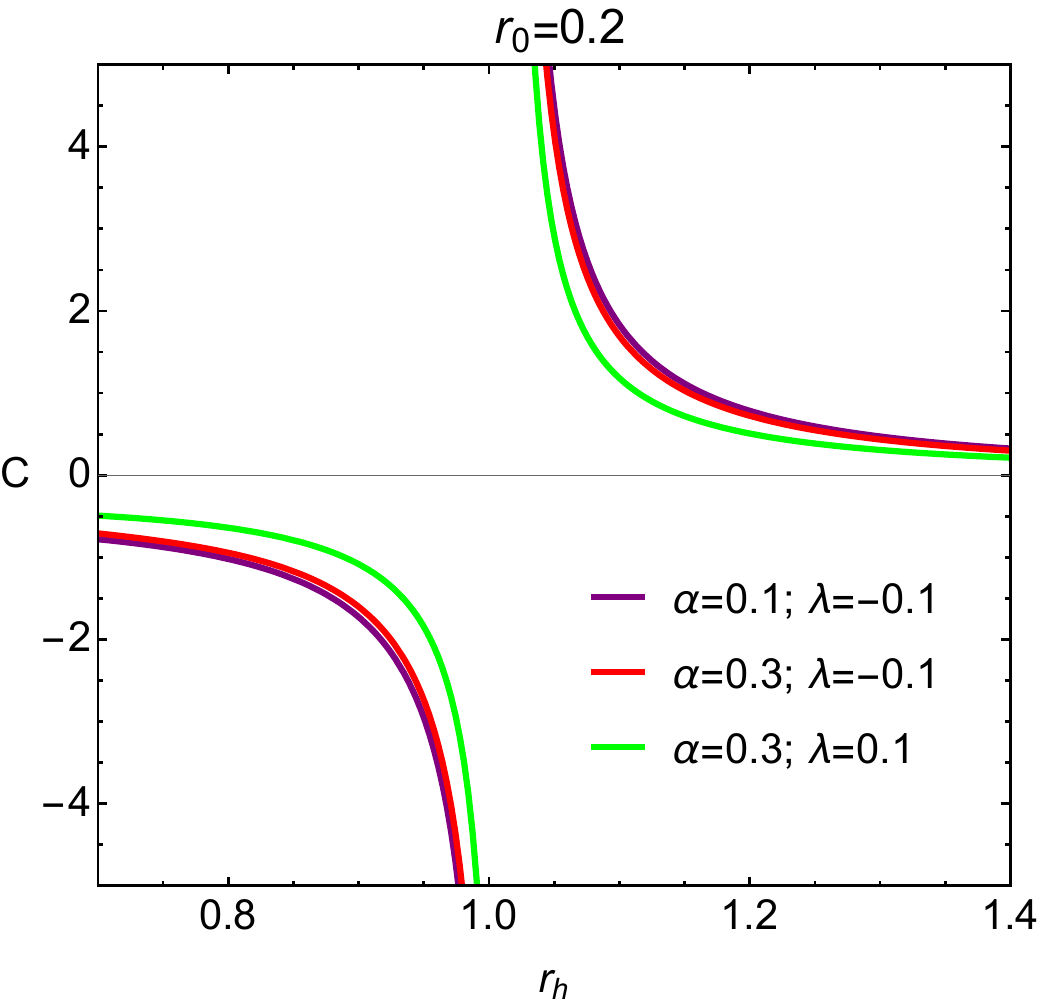}
    \caption{Heat capacity $C$ as a function of the horizon radius $r_h$ for two values of $r_0$. The divergence of $C$ indicates a critical point associated with a phase transition. The parameters $\alpha$ and $\lambda$ shift the location of this critical radius and modify the thermodynamic stability region of the black hole.}
    \label{specific}
\end{figure}

In Fig \ref{specific}, the heat capacity $C$ is presented as a function of the horizon radius $r_h$, revealing the thermodynamic stability structure of the black hole. The vertical divergence observed in each panel corresponds to a critical radius at which $C$ diverges, indicating the occurrence of a phase transition. At this point, the Hawking temperature reaches an extremum with respect to $r_h$, and the heat capacity changes its sign. In the region preceding the divergence, $C<0$, characterizing thermodynamically unstable configurations, whereas beyond the critical point $C>0$, corresponding to a stable black hole phase.
The role of the parameters is clearly manifested in the shift of the critical radius and in the magnitude of the stable branch. An increase in $\alpha$ modifies the geometric structure of the spacetime and slightly displaces the transition point, while also affecting the growth of the positive heat capacity region. The parameter $\lambda$, associated with exotic matter (PFDM), produces a more significant shift of the divergence point, demonstrating its strong influence on the phase structure and stability domain of the black hole.
Furthermore, the comparison between the left and right panels shows that the parameter $r_0$ rescales the thermodynamic behavior. For larger $r_0$, the stable region extends over a wider range of $r_h$, and the positive branch of $C$ becomes more pronounced. In the large $r_h$ limit, all curves gradually converge toward zero, indicating that the impact of the additional model parameters weakens at large scales and the system approaches the classical black hole regime.

From the above analysis, it is clear that the thermodynamic properties include the Hawking temperature and the specific heat capacity are influenced by the geometric parameters ($r_0,\,r_g,\,\lambda,\,\alpha$) that alters the space-time curvature. Consequently, the result gets modification in comparison to the Dymnikova black hole.  

\section{Dynamics of Massless Particles}\label{sec:4}

In this section, we study the geodesic motion of massless particles (photons) around the Dymnikova black hole, taking into account the presence of perfect fluid dark matter (PFDM) and a cloud of strings. We focus on key features of the photon dynamics, including the photon sphere, the black hole shadow, the effective radial force, and the photon trajectories. In particular, we analyze how the geometric parameters $(r_0,\,\alpha,\,\lambda)$ affect these optical properties. A lot of studies on photon dynamics have been conducted in various black holes configurations both in general relativity and modified gravity theories (see \cite{FA1,FA2,FA3,FA4,FA5,FA6}) and therein related references.

The space-time (\ref{metric}) can be expressed as $ds^2=g_{\mu\nu} dx^{\mu} dx^{\nu}$, where the metric tensor $g_{\mu\nu}$ with $\mu,\nu=0,...,3$ is given by
\begin{equation}
    g_{\mu\nu}=\mbox{diag}\left(-f(r),\,\frac{1}{f(r)},\,r^2,\,r^2 \sin^2 \theta\right),\label{bb1} 
\end{equation}

To study null geodesics, we approach the Lagrangian formalism through which we derive the effective that governs the dynamics of massless particles. According to this formalism, the Lagrangian density function in terms of the metric tensor $g_{\mu\nu}$ can be written as
\begin{equation}
    \mathbb{L}=\frac{1}{2}g_{\mu\nu} \dot x^{\mu}\,\dot x^{\nu},\label{bb2}
\end{equation}
where dot represents ordinary derivative w. r. to $\lambda$, an affine parameter along geodesics.

Considering the geodesic motion on the equatorial plane defined by $\theta=\pi/2$, the Lagrangian density function (\ref{bb2}) using (\ref{bb1}) explicitly can be written as
\begin{equation}
    \mathbb{L}=\frac{1}{2}\left[-f(r)\,\dot t^2+\frac{1}{f(r)} \dot r^2+r^2 \dot \phi^2 \right].\label{bb3}
\end{equation}

The Lagrangian density function is independent of the temporal coordinate $t$ and the angular coordinate $\theta$. Therefore, there exist two conserved quantities called the conserved energy and conserved angular momentum. These are given by
\begin{equation}
    \mathrm{E}=f(r)\,\dot t,\qquad \mathrm{L}=r^2\,\dot \phi.\label{bb4}
\end{equation}

For photon motions, $g_{\mu\nu} \dot x^{\mu}\,\dot x^{\nu}=0$ implies the following equation:
\begin{equation}
    \dot r^2+V_{\rm eff}=\mathrm{E}^2,\label{bb6}
\end{equation}
where $V_{\rm eff}$ is the effective potential governing the dynamics of massless particles and is given by
\begin{equation}
    V_{\rm eff}=\frac{\mathrm{L}^2}{r^2}\,f(r)=\frac{\mathrm{L}^2}{r^2}\,\left[1-\alpha - \dfrac{r_g}{r}\left(1 - e^{-r^{3}/r_{*}^3}\right)+\dfrac{\lambda}{r}\ln\!\dfrac{r}{|\lambda|}\right].\label{bb7}
\end{equation}

\begin{figure}[ht!]
    \centering
    \includegraphics[width=0.45\linewidth]{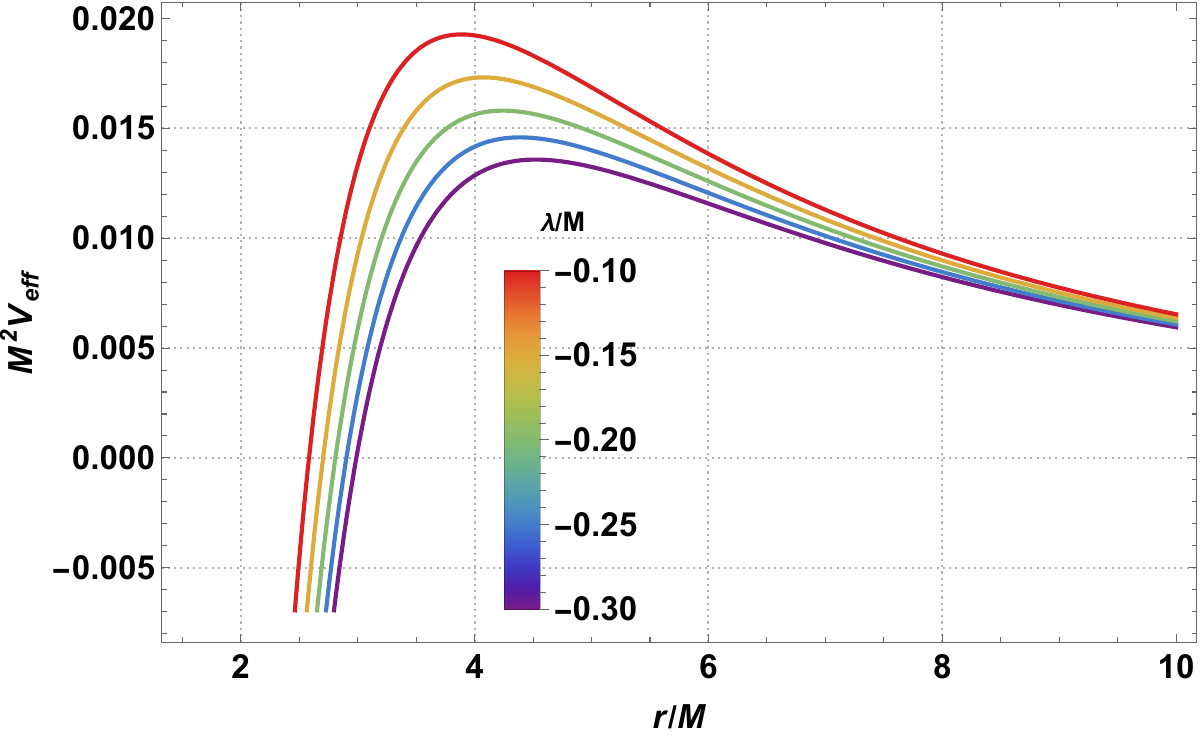}\qquad
    \includegraphics[width=0.45\linewidth]{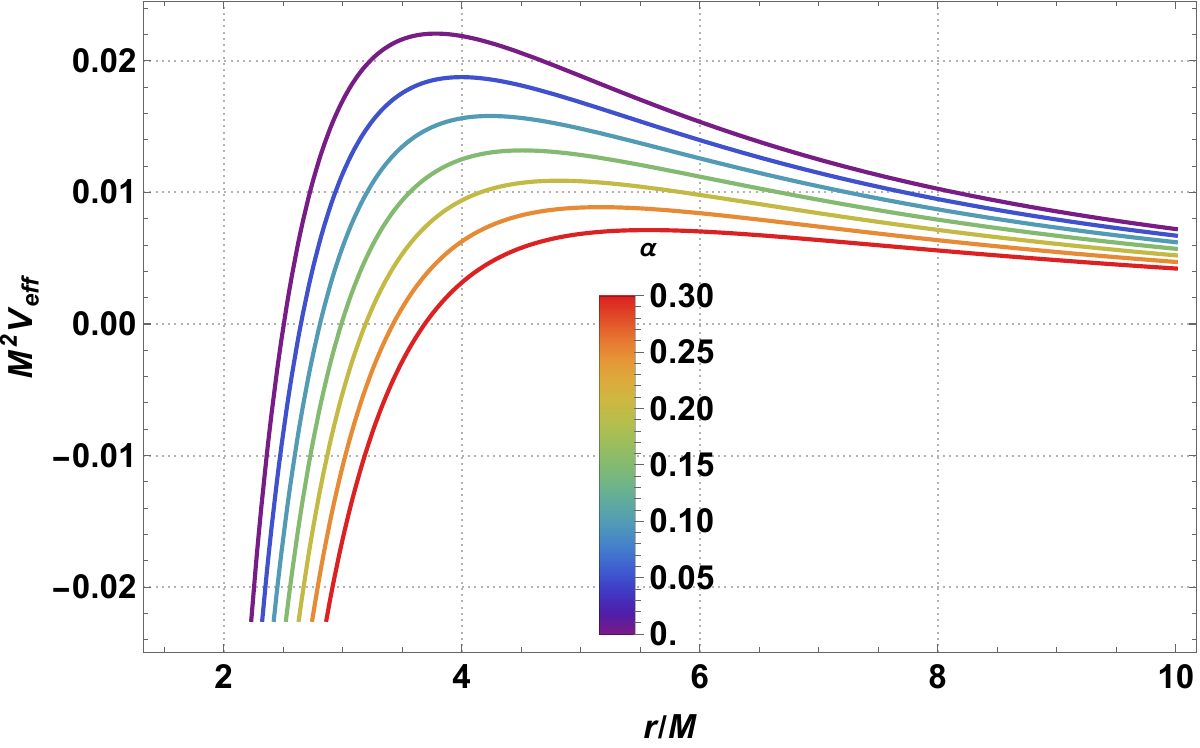}
    \caption{The behavior of the effective potential as a function of radial distance by varying $\lambda$ and $\alpha$.}
    \label{fig:null}
\end{figure}

Figure \ref{fig:null} shows the effective potential $V_{\rm eff}$ as a function of the dimensionless radial coordinate $r/M$, illustrating the characteristic barrier structure associated with circular motion in strong gravitational fields. The presence of a local maximum determines the stability conditions of particle orbits and governs the location of marginally stable configurations.
Varying the PFDM parameter $\lambda$ significantly modifies both the height and the radial position of the potential peak. As $\lambda$ becomes more negative, the maximum of $V_{\rm eff}$ increases and shifts outward, indicating a strengthening of the effective gravitational attraction and a deeper confinement region for test particles. This directly affects the location of stable circular orbits and alters the epicyclic frequencies responsible for QPO formation.
Increasing the deformation parameter $\alpha$ systematically lowers the potential barrier and shifts its position, reflecting modifications of spacetime curvature in the strong–field regime. A larger $\alpha$ reduces the height of the effective potential, implying that particles require lower energy to approach the central region. Consequently, the positions of marginally stable and marginally bound orbits are displaced, leading to measurable changes in the orbital and resonance frequency spectrum.
At large $r/M$, all curves converge, confirming that PFDM and deformation effects become negligible in the weak–field limit and the potential approaches the classical behavior. In contrast, within the intermediate region $r/M \sim 3$–6, the deviations are substantial, indicating that environmental and geometric corrections strongly influence the stability structure and resonance radii relevant for QPO generation.

For circular null orbits of fixed radii $r=r_c$, the conditions $\dot r=0$ and $\ddot r=0$ must be satisfied. From Eq. (\ref{bb6}), we find the following relations:
\begin{equation}
    \mathrm{E}^2=V_{\rm eff}=\frac{\mathrm{L}^2}{r^2}\,f(r).\label{bb8}
\end{equation}
And
\begin{equation}
    \frac{d}{dr}\left(\frac{f(r)}{r^2}\right)\Big{|}_{r=r_s}=0\Rightarrow 2 f(r_s)-r_s\,f'(r_s)=0.\label{bb9}
\end{equation}

The relation (\ref{bb9}) gives us the photon sphere radius $r=r_s$ satisfying the following polynomial equation as,
\begin{equation}
1-\alpha- \frac{3 r_g}{2r_s} \left[1 -\left(1+\frac{r^3_s}{r_*^{3}}\right)\,e^{-r^3_s/r_*^3}\right]
+ \frac{\lambda}{2r_s} \left(3\ln\!\frac{r_s}{|\lambda|} -1\right)= 0.\label{bb10}
\end{equation}
The exact analytical solution of the above polynomial represents the photon sphere radius $r_s$. Noted that the exact analytical solution is quite a challenging due to the presence of both the logarithmic and exponential term. However, one can determine the numerical results by selecting suitable values of string cloud parameter $\alpha$, the de Sitter radius $r_0$, and PFDM parameter $\lambda$.

To determine the shadow radius, we examine the behavior of the metric function $f(r)$ at large distances. Noted that 
\begin{equation}
    \lim_{r \to \infty} f(r) = 1 - \alpha \neq 1,\label{bb11}
\end{equation}
which indicates that the function is asymptotically bounded rather than flat space. Therefore, the shadow radius for a static observer located at position $r_{\rm obs}$ is defined by \cite{Volker2022}
\begin{equation}
    R_{\rm sh} \simeq r_{\rm obs} \theta_{\rm sh}=r_s \sqrt{\frac{f(r_{\rm obs})}{f(r_s)}}=r_s \sqrt{\frac{1-\alpha - \dfrac{r_g}{r_{\rm obs}}\left(1 - e^{-r_{\rm obs}^{3}/r_{*}^3}\right)+\dfrac{\lambda}{r_{\rm obs}}\ln\!\dfrac{r_{\rm obs}}{|\lambda|}}{1-\alpha - \dfrac{r_g}{r_s}\left(1 - e^{-r_s^{3}/r_{*}^3}\right)+\dfrac{\lambda}{r_s}\ln\!\dfrac{r_s}{|\lambda|}}}.\label{bb12}
\end{equation}
For a distant observer ($r_{\rm obs} \to \infty$), the shadow radius simplifies as,
\begin{equation}
    R_{\rm sh}=(1-\alpha)^{1/2}\frac{r_s }{\sqrt{1-\alpha - \dfrac{r_g}{r_s}\left(1 - e^{-r_s^{3}/r_{*}^3}\right)+\dfrac{\lambda}{r_s}\ln\!\dfrac{r_s}{|\lambda|}}}=(1-\alpha)^{1/2}\,\beta_c,\label{bb13}
\end{equation}
where $\beta_c=\mathrm{L}/\mathrm{E}$ is the critical impact parameter of massless particles that can be determined using the relation (\ref{bb8}) at radius $r=r_s$.

\begin{table}[ht]
\centering
\begin{tabular}{|c| c| c| c|}
\hline
$\alpha$ & $\lambda/M$ & $r_s/M$ & $R_{\rm Sh}/M$ \\
\hline
0.05 & -0.3 & 4.2564  & 7.3714  \\
0.05 & -0.2 & 3.99854 & 6.87908 \\
0.05 & -0.1 & 3.67431 & 6.27654 \\
0.05 &  0.1 & 2.69068 & 4.50499 \\
0.05 &  0.2 & 2.46944 & 4.06687 \\
\hline
0.10 & -0.3 & 4.52327 & 7.81396 \\
0.10 & -0.2 & 4.24024 & 7.27979 \\
0.10 & -0.1 & 3.88785 & 6.63030 \\
0.10 &  0.1 & 2.83165 & 4.73681 \\
0.10 &  0.2 & 2.59066 & 4.26319 \\
\hline
0.15 & -0.3 & 4.82335 & 8.30911 \\
0.15 & -0.2 & 4.51156 & 7.72766 \\
0.15 & -0.1 & 4.12709 & 7.02521 \\
0.15 &  0.1 & 2.98869 & 4.99456 \\
0.15 &  0.2 & 2.72519 & 4.48073 \\
\hline
0.20 & -0.3 & 5.16310 & 8.86654 \\
0.20 & -0.2 & 4.81819 & 8.23137 \\
0.20 & -0.1 & 4.39690 & 7.46881 \\
0.20 &  0.1 & 3.16475 & 5.28291 \\
0.20 &  0.2 & 2.87539 & 4.72320 \\
\hline
0.25 & -0.3 & 5.55074 & 9.49847 \\
0.25 & -0.2 & 5.16739 & 8.80185 \\
0.25 & -0.1 & 4.70351 & 7.97060 \\
0.25 &  0.1 & 3.36355 & 5.60769 \\
0.25 &  0.2 & 3.04426 & 4.99526 \\
\hline
0.30 & -0.3 & 5.99693 & 10.2205 \\
0.30 & -0.2 & 5.56853 & 9.45303 \\
0.30 & -0.1 & 5.05492 & 8.54269 \\
0.30 &  0.1 & 3.58985 & 5.97637 \\
0.30 &  0.2 & 3.23558 & 5.30284 \\
\hline
\end{tabular}
\caption{Photon sphere radius $r_s$ and shadow radius $R_{\rm Sh}$ for different values of $\alpha$ and $\lambda$. Here $r_0/M=0.4,\,r_g/M=2,\,r_{\rm obs}/M=50$.}
\label{tab:1}
\end{table}

\begin{figure}[ht!]
    \centering
    \includegraphics[width=0.45\linewidth]{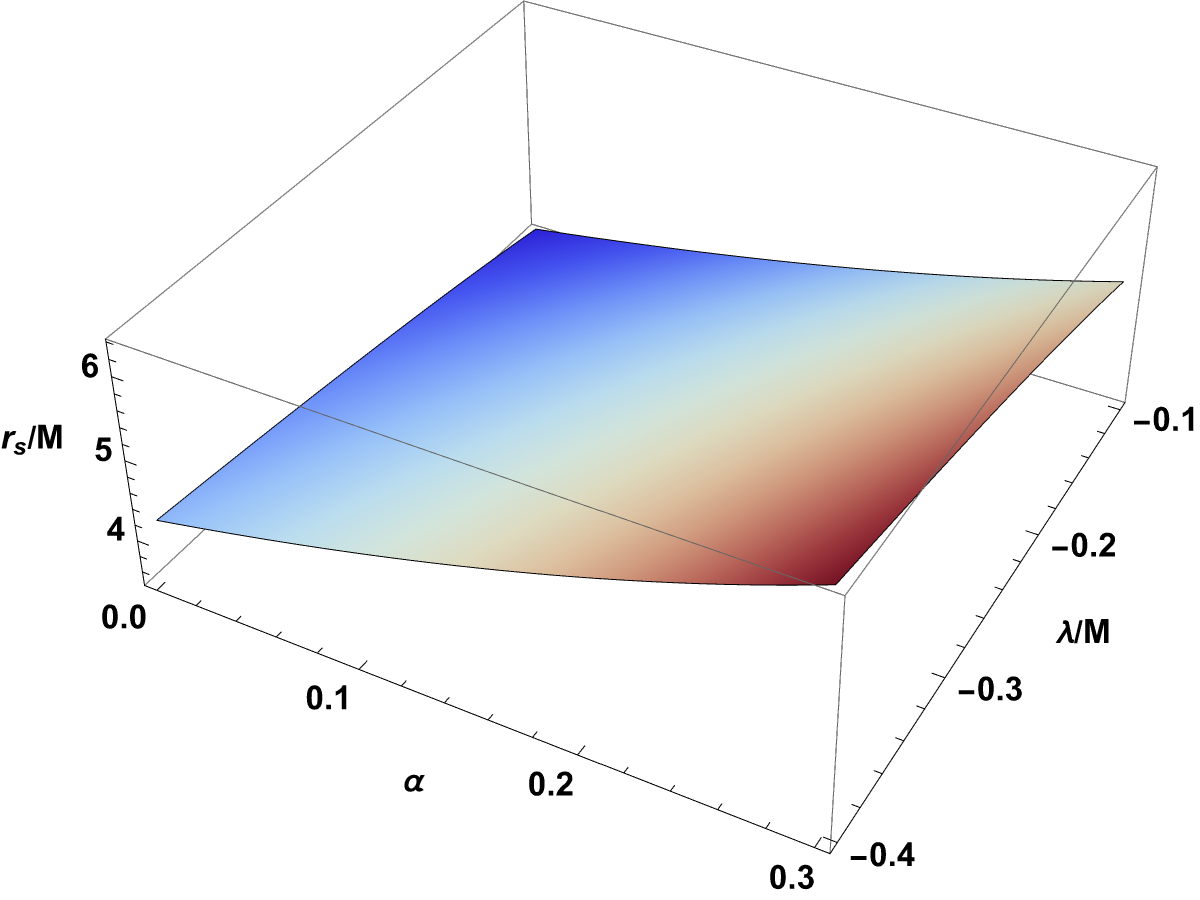}\qquad
    \includegraphics[width=0.45\linewidth]{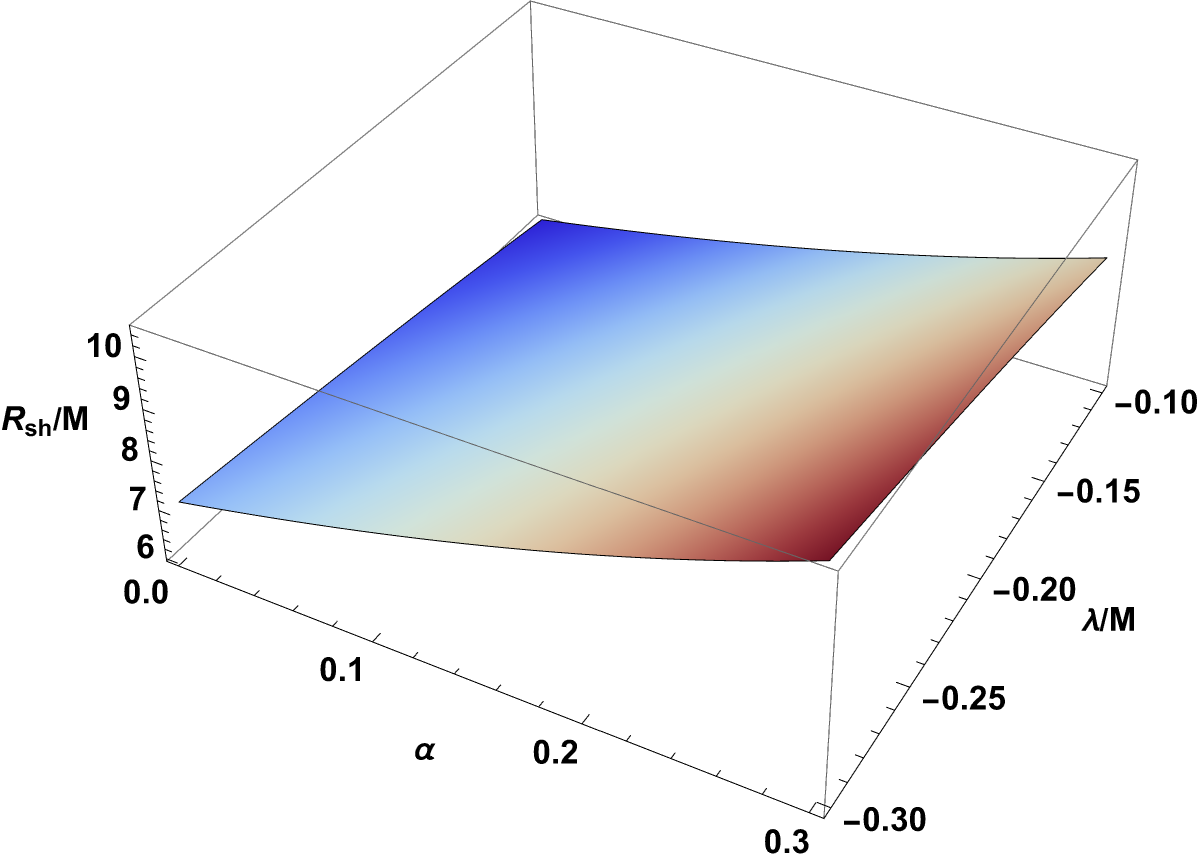}
    \caption{Photon sphere and shadow radii. Here $r_0=0.4,\, r_{\rm obs}/M=100$.}
    \label{fig:photon-shadow}
\end{figure}

From the analysis of photon dynamics, it is clear that the photon sphere and the black hole shadow are influenced by the geometric parameters ($r_0,\,\lambda,\,\alpha$). Consequently, the result gets modification in comparison to the Dymnikova black hole case. In table \ref{tab:1}, we presented numerical values of the photon sphere and shadow radii by varying together the PFDM parameter $\lambda$ and string cloud parameter $\alpha$.

\section{Dynamics of Massive Test Particle}\label{sec:5}

In this section, we study the geodesic motion of massive test particles around the Dymnikova black hole, taking into account the presence of perfect fluid dark matter (PFDM) and a cloud of strings. We focus on key features of the particle dynamics, including the specific energy, specific angular momentum, innermost stable circular orbits (ISCO), and the effective force experiences by this. In particular, we analyze how the geometric parameters $(r_0,\,\alpha,\,\lambda)$ affect these dynamical characteristics. A lot of studies have been conducted the dynamics of test particles around various black hole configurations in the literature (see, for example, \cite{FA1,FA2,FA3,FA4,FA5,FA6}).

The Lagrangian density function of massive particles of mass $m$ can be written as
\begin{equation}
    \mathbb{L}=\frac{1}{2}\,m\,g_{\mu\nu}\,\frac{dx^{\mu}}{d\tau}\,\frac{dx^{\nu}}{d\tau},\label{lag}
\end{equation}

Following the Euler-Lagrange method, we find 
\begin{eqnarray}
p_{t}/m &=& -\left(1-\alpha - \dfrac{r_g}{r}\left(1 - e^{-r^{3}/r_{*}^3}\right)+\dfrac{\lambda}{r}\ln\!\dfrac{r}{|\lambda|}\right)\,\frac{dt}{d\tau}=-\mathcal{E},\label{bb16}\\
p_{\phi}/m &=& r^{2}\sin^{2}\theta \,\frac{d\phi}{d\tau}=\mathcal{L}\, , \label{bb17} \\
p_{r}/m &=& \left(1-\alpha - \dfrac{r_g}{r}\left(1 - e^{-r^{3}/r_{*}^3}\right)+\dfrac{\lambda}{r}\ln\!\dfrac{r}{|\lambda|}\right)^{-1}\,\frac{dr}{d\tau}, \label{bb18}\\
p_{\theta}/m &=& r^{2}\,\frac{d\theta}{d\tau},\label{bb19}
\end{eqnarray}
where $\mathcal{L}$ and $\mathcal{E}$ are the orbital angular momentum and the energy per unit mass of the test particle, respectively.

The first integral equations of motion therefore are given by
\begin{align}
&\frac{dt}{d\tau}=\mathcal{E}\,\left(1-\alpha - \dfrac{r_g}{r}\left(1 - e^{-r^{3}/r_{*}^3}\right)+\dfrac{\lambda}{r}\ln\!\dfrac{r}{|\lambda|}\right)^{-1},\nonumber\\
&\frac{d\phi}{d\tau}=\mathcal{L}/(r^2\,\sin^2 \theta),\nonumber\\
&\frac{d\theta}{d\tau}=p_{\theta}/(m\,r^2).\label{integral}
\end{align}

Employing the normalization condition, $g_{\mu\nu}\dot{x}^{\mu}\dot{x}^{\nu}=-1$ for massive test particles, the equation of motion for the radial coordinate $r$ is given by
\begin{eqnarray}
\left(\frac{dr}{d\tau}\right)^{2}+\frac{p^2_{\theta}}{m^2\,r^2}\,\left(1-\alpha - \dfrac{r_g}{r}\left(1 - e^{-r^{3}/r_{*}^3}\right)+\dfrac{\lambda}{r}\ln\!\dfrac{r}{|\lambda|}\right)=U_{\rm eff}\, ,\label{bb20}
\end{eqnarray}
where the effective potential $U_{\rm eff}$ of the system is given by
\begin{equation}
U_{\rm eff}=\mathcal{E}^{2}-\left(1-\alpha - \dfrac{r_g}{r}\left(1 - e^{-r^{3}/r_{*}^3}\right)+\dfrac{\lambda}{r}\ln\!\dfrac{r}{|\lambda|}\right)\left(1+\frac{\mathcal{L}^{2}}{r^{2}\,\sin^2 \theta}\right)\, .\label{bb21}
\end{equation}
One can see that the effective potential governing the particle dynamics is influenced by the geometric parameters. These include the string cloud parameter $\alpha$, PFDM parameter $\lambda$, de Sitter radius $r_0$. Moreover, the angular momentum per unit mass $\mathcal{L}$ also alter the effective potential.

For circular motion of test particles on the equatorial plane ($\theta=\pi/2,\,p_{\theta}=0$) of fixed radius, the conditions $\dot r=0$ and $\ddot r=0$ must be satisfied. Using Eq.~(\ref{bb20}), we find
\begin{equation}
U_{\rm eff}=0 \Rightarrow \mathcal{E}^{2}=\left(1-\alpha - \dfrac{r_g}{r}\left(1 - e^{-r^{3}/r_{*}^3}\right)+\dfrac{\lambda}{r}\ln\!\dfrac{r}{|\lambda|}\right)\left(1+\frac{\mathcal{L}^{2}}{r^{2}}\right).\label{bb22}
\end{equation}
And
\begin{equation}
    \frac{\partial U_{\rm eff}}{\partial r}\Big{|}_{r=\mbox{const.}}=0.\label{bb23}
\end{equation}

Substituting the potential $U_{\rm eff}$ given in (\ref{bb20}) into the condition (\ref{bb23}) and after simplification results
\begin{align}
\mathcal{L}_{\rm sp}=r\,\sqrt{\frac{\frac{r_g}{2r}\left\{1-\left(1+\frac{3 r^3}{r^3_*}\right)\,e^{-r^{3}/r_*^{3}}\right\}+\frac{\lambda}{2r}\left(1-\ln\!\frac{r}{|\lambda|}\right)}{1-\alpha- \frac{3 r_g}{2r} \left\{1 -\left(1+\frac{r^3}{r_*^{3}}\right)\,e^{-r^3/r_*^3}\right\}
+ \frac{\lambda}{2r} \left(3\ln\!\frac{r}{|\lambda|} -1\right)}}\Bigg{|}_{r=\mbox{const.}}.\label{bb24}
\end{align}
Using $\mathcal{L}_{\rm sp}$ into the relation (\ref{bb22}) yields
\begin{align}
\mathcal{E}_{\rm sp}=\frac{\left(1-\alpha - \dfrac{r_g}{r}\left(1 - e^{-r^{3}/r_{*}^3}\right)+\dfrac{\lambda}{r}\ln\!\dfrac{r}{|\lambda|}\right)}{\sqrt{1-\alpha- \frac{3 r_g}{2r} \left\{1 -\left(1+\frac{r^3}{r_*^{3}}\right)\,e^{-r^3/r_*^3}\right\}
+ \frac{\lambda}{2r} \left(3\ln\!\frac{r}{|\lambda|} -1\right)}}\Bigg{|}_{r=\mbox{const.}}.\label{bb25}
\end{align}
Here $\mathcal{L}_{\rm sp}$ is the specific angular momentum and $\mathcal{E}_{\rm sp}$ is the specific energy associated with massive test particles orbiting around the black hole in fixed radius.

It is observed that the specific energy and angular momentum of test particles undergoing circular motion depend sensitively on the string cloud parameter $\alpha$, the PFDM parameter $\lambda$, and the de Sitter radius $r_0$.

\begin{figure}[ht!]
    \centering
    \includegraphics[width=0.45\linewidth]{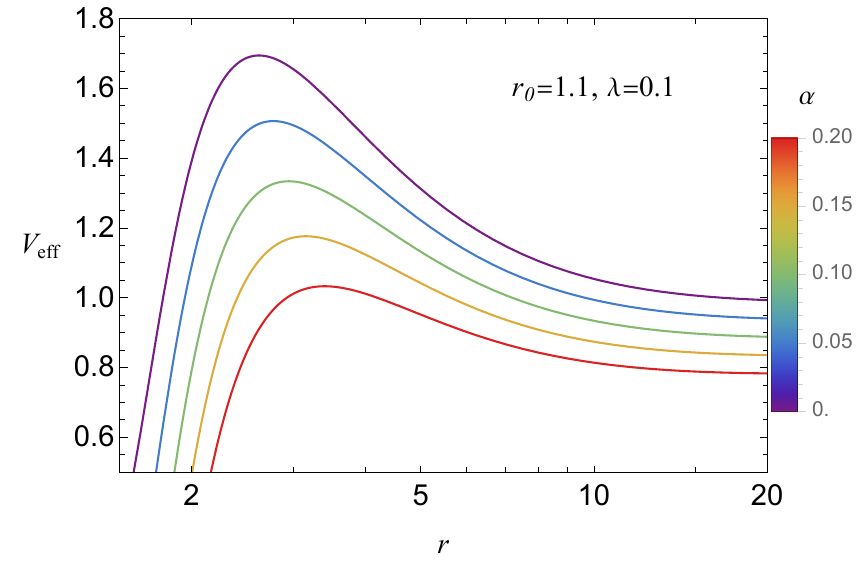}
    \includegraphics[width=0.45\linewidth]{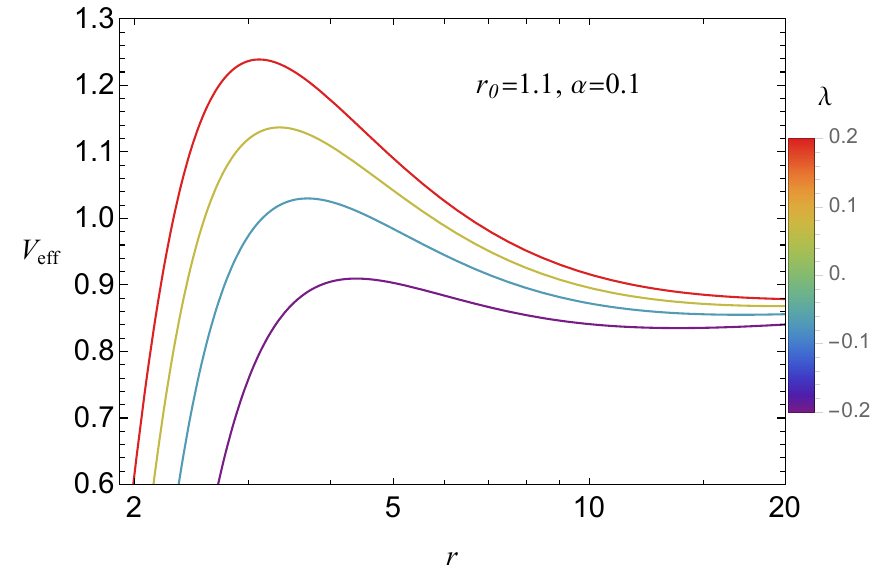}
    \includegraphics[width=0.45\linewidth]{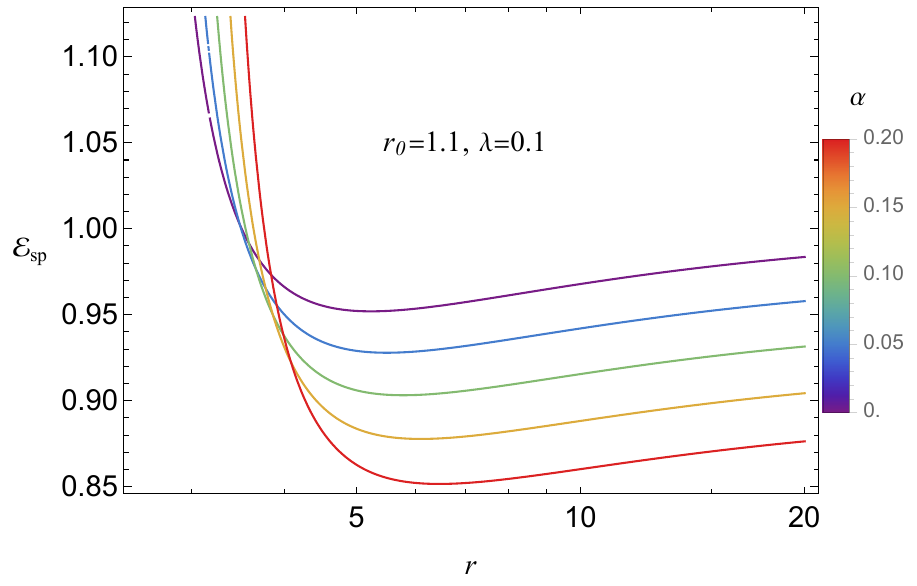}
    \includegraphics[width=0.45\linewidth]{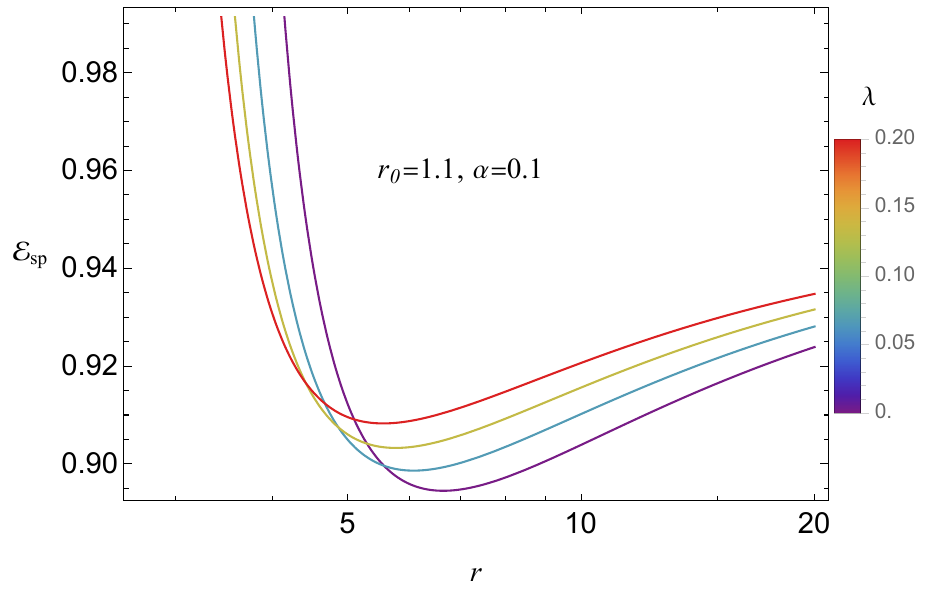}
    \includegraphics[width=0.45\linewidth]{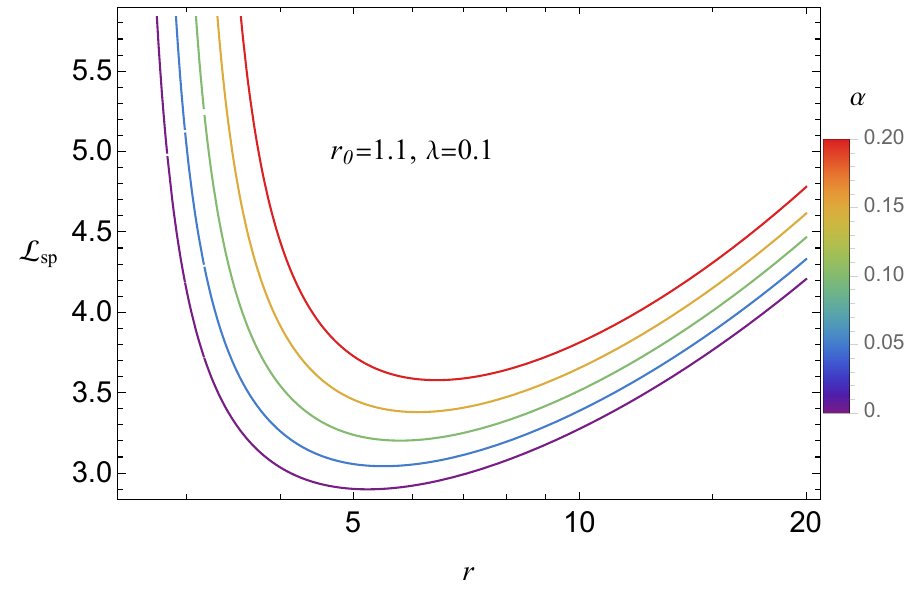}
    \includegraphics[width=0.45\linewidth]{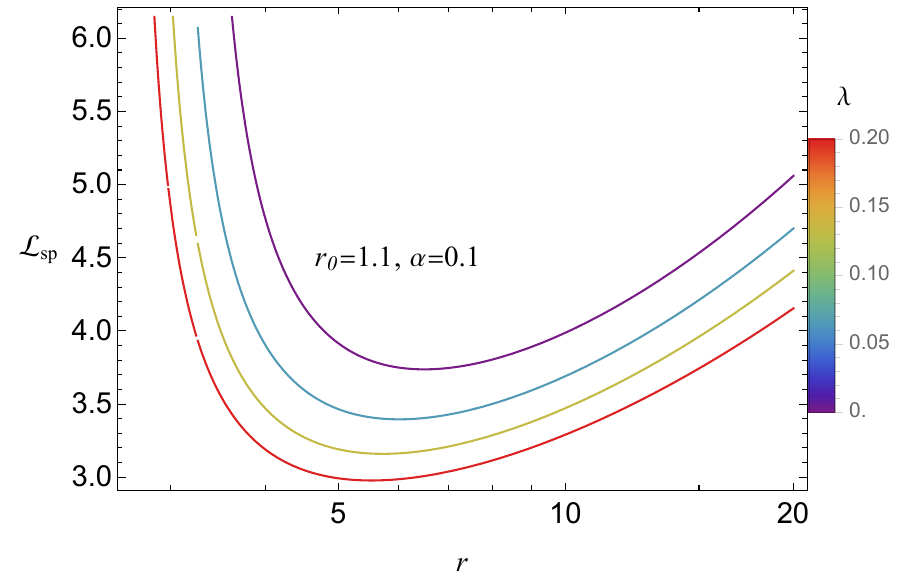}
    \caption{Radial behaviour of the effective potential $V_{\rm eff}$  together with the specific energy ${\cal E}_{sp}$ and the specific angular momentum ${\cal L}_{sp}$. In the left column, $\lambda$ is kept fixed while $\alpha$ is varied. In the right column, $\alpha$ is kept fixed while $\lambda$ is varied. All curves are shown as functions of the radial coordinate $r$.}
    \label{VLE}
\end{figure}

Figure~\ref{VLE} demonstrates that the orbital dynamics is primarily governed by the shape and amplitude of the effective potential $V_{\rm eff}$ . The extrema of $V_{\rm eff}$  encode the existence of circular configurations and their stability character: when the effective barrier is enhanced, access to the inner region becomes dynamically suppressed, whereas a reduced barrier favours inward motion and modifies the separation between capture and scattering regimes. The trends in the panels indicate that increasing $\alpha$ tends to lower $V_{\rm eff}$  and soften the effective barrier, while increasing $\lambda$ generally strengthens the potential profile and makes the barrier more pronounced. Consequently, the characteristic  radii associated with the transition between different dynamical regimes shift systematically as the parameters are varied.

The behaviour of the specific energy ${\cal E}_{sp}$ and the specific angular momentum ${\cal L}_{sp}$ provides complementary information by quantifying the requirements for sustaining circular motion at a given radius. Their characteristic minima signal the approach to a marginal configuration and are therefore directly relevant for identifying the boundary of the stable-orbit domain. The curves show that increasing $\alpha$ typically lowers ${\cal E}_{sp}$, indicating more tightly bound circular configurations, but raises the required ${\cal L}_{sp}$, implying a larger angular-momentum demand to maintain circular motion. In contrast, increasing $\lambda$ tends to increase ${\cal E}_{sp}$ while reducing ${\cal L}_{sp}$, circular motion becomes energetically more demanding but requires less angular momentum. Overall, the competing impact of $\alpha$ and $\lambda$ reshapes the domain of stable circular motion and shifts the critical radii controlling the inner edge of circular configurations, which is expected to translate into measurable changes in disk-related orbital properties.

Moreover, stable circular orbits satisfy the following conditions:
\begin{eqnarray}
U_{\rm eff}=0,\qquad \frac{\partial U_{\rm eff}}{\partial r}=0,\qquad \frac{\partial^2 U_{\rm eff}}{\partial r^2} \geq 0.\label{bb26}
\end{eqnarray}
For a marginally stable circular orbits, we must have $\frac{\partial^2 U_{\rm eff}}{\partial r^2}=0$. Thereby, using the potential in (\ref{bb21}), we find the following compact relation in terms of the lapse function $f$ as,
\begin{equation}
    \frac{3}{r}\,f(r)\,f'(r)-2 (f'(r))^2+f(r)\,f''(r)=0.\label{bb27}
\end{equation}
Substituting the selected metric function $f(r)$ into the above relation (\ref{bb27}) results a higher polynomial relation in $r$ whose exact analytical solution will give us the innermost stable circular orbits (ISCO) radius. Noted that exact analytical solution of the polynomial equation is quite a challenging task due to the presence of exponential term. However, one can attempt to find numerical values of ISCO radius by selecting suitable values of the geometric parameters $(r_0,\,\alpha,\,\lambda)$.

\begin{figure}[ht!]
    \centering
    \includegraphics[width=0.55\linewidth]{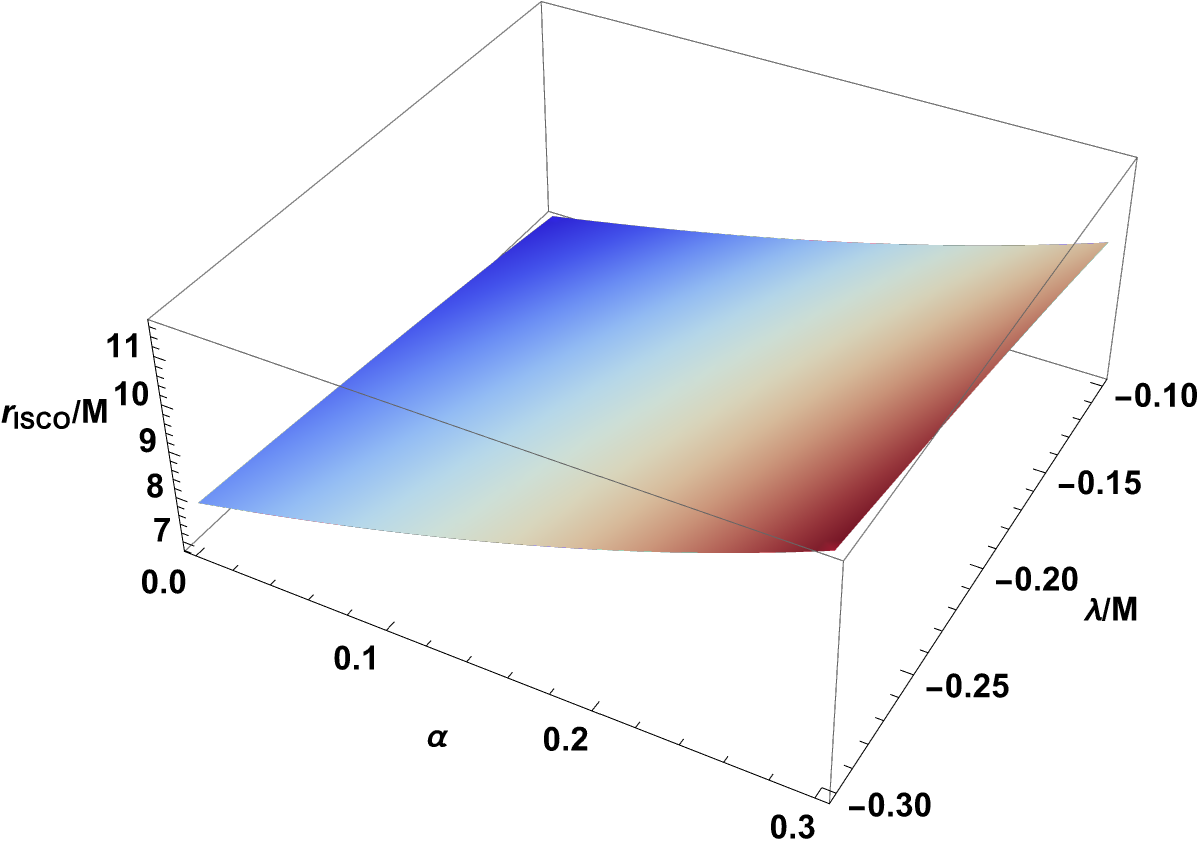}
    \caption{Behavior of the ISCO radius. Here $r_0/M=0.1$.}
    \label{fig:ISCO}
\end{figure}

\begin{figure}[ht!]
    \centering
    \includegraphics[width=0.45\linewidth]{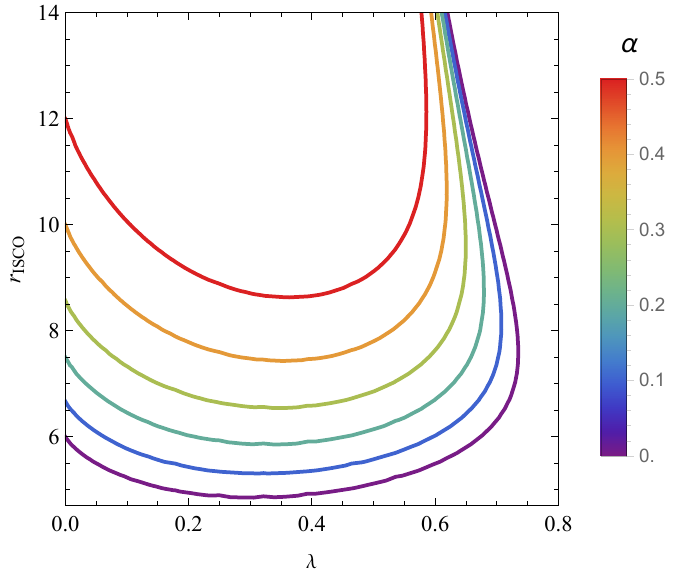}
    \includegraphics[width=0.45\linewidth]{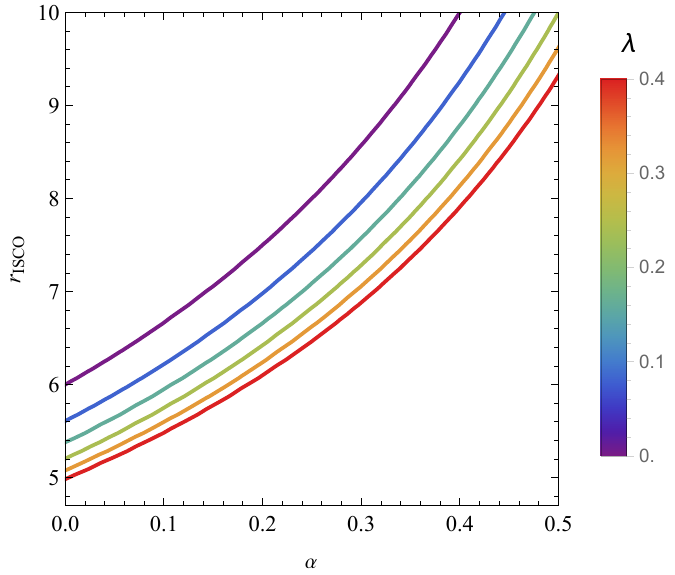}
    \caption{Dependence of the radius of the innermost stable circular orbit $r_{\mathrm{ISCO}}$ on the model parameters. The left panel shows the variation of $r_{\mathrm{ISCO}}$ with the PFDM parameter $\lambda$ for several values of the deformation parameter $\alpha$, while the right panel presents the dependence on $\alpha$ for different $\lambda$. In all calculations the scale parameter is fixed at $r_0 = 1.2$.}
    \label{risco}
\end{figure}

Figure \ref{risco} demonstrates how the location of the innermost stable circular orbit is modified by the combined influence of spacetime deformation and the surrounding matter distribution. 
Since $r_{\mathrm{ISCO}}$ directly traces the strength of the effective gravitational potential, an inward shift corresponds to stronger binding, while an outward displacement indicates earlier instability and weaker confinement. 
These variations determine the position of the inner accretion flow and therefore affect the dynamical properties of matter near the compact object.

In the left panel, the dependence on the PFDM parameter is clearly non–monotonic. For small $\lambda$, the orbit moves inward and reaches a minimum, implying a deepening of the potential well and allowing particles to remain stable closer to the center. Beyond this region, further increase of $\lambda$ drives the orbit outward, signaling that the environmental contribution begins to screen gravity or destabilize circular motion. Larger values of $\alpha$ systematically shift the curves toward greater radii, showing that stronger deviations from the classical geometry weaken the effective attraction.

The right panel emphasizes the global role of the deformation parameter. For any chosen $\lambda$, $r_{\mathrm{ISCO}}$ increases steadily with $\alpha$, meaning that stability can only be preserved at larger distances as the spacetime departs from the standard limit. At the same time, higher $\lambda$ keeps the orbit closer to the compact object, indicating that the surrounding matter partially compensates the destabilizing effect of the deformation. 
The balance between these two contributions ultimately fixes the inner boundary of stable motion.

\section{Quasi-periodic Oscillations (QPOs)}\label{sec:6}

In this subsection, we analyze the expected behavior of the lower and upper QPO frequencies within several theoretical frameworks commonly used to interpret twin-peak structures. These models relate the observed signals to combinations of the fundamental orbital and epicyclic frequencies of test particles moving in the gravitational field of compact objects.

\begin{itemize}

\item The relativistic precession (RP) model, originally proposed by Stella and Vietri, was developed to account for kHz twin-peak QPOs detected in low-mass X-ray binaries within the approximate range $0.2$--$1.25$ kHz. The same formalism can be extended to black hole systems, where timing properties of the accretion flow enable constraints on the mass and spin of the central object through analysis of the power-density spectrum. In this scenario, the characteristic frequencies are directly associated with circular motion near the ISCO, with the identification $\nu_U = \nu_\phi$ and $\nu_L = \nu_\phi - \nu_r$ \cite{StellaVietri1998,StellaVietriMorsink1999,StellaVietriPRL1999}.

\item The warped disk (WD) model is based on non-axisymmetric oscillatory behavior of matter in thin disks around compact objects. In this framework, QPO frequencies emerge from deformations that introduce vertical motion and lead to a warped disk configuration. The corresponding relations between the upper and lower frequencies are given by $\nu_U = 2\nu_\phi - \nu_r$ and $\nu_L = 2(\nu_\phi - \nu_r)$ \cite{Kato2004Warp,Kato2008Deformed,Kato2007Corr}.
\end{itemize}

\begin{figure*}
    \centering
    \includegraphics[width=0.49\linewidth]{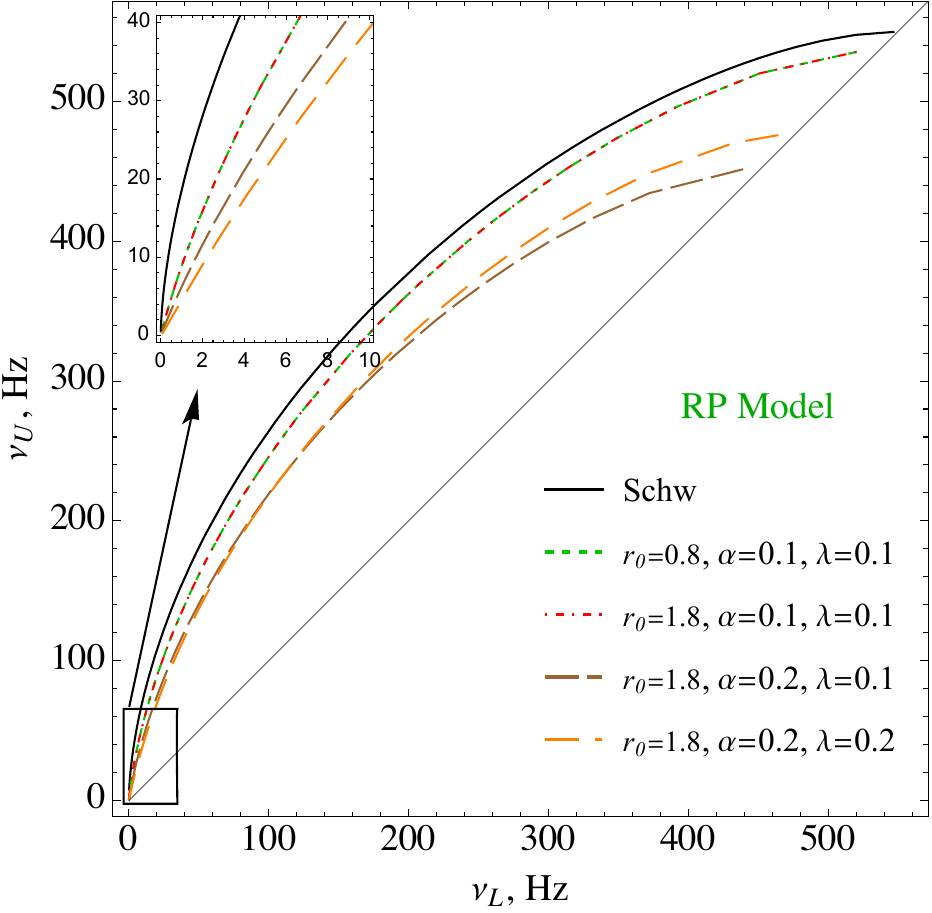}
    \includegraphics[width=0.49\linewidth]{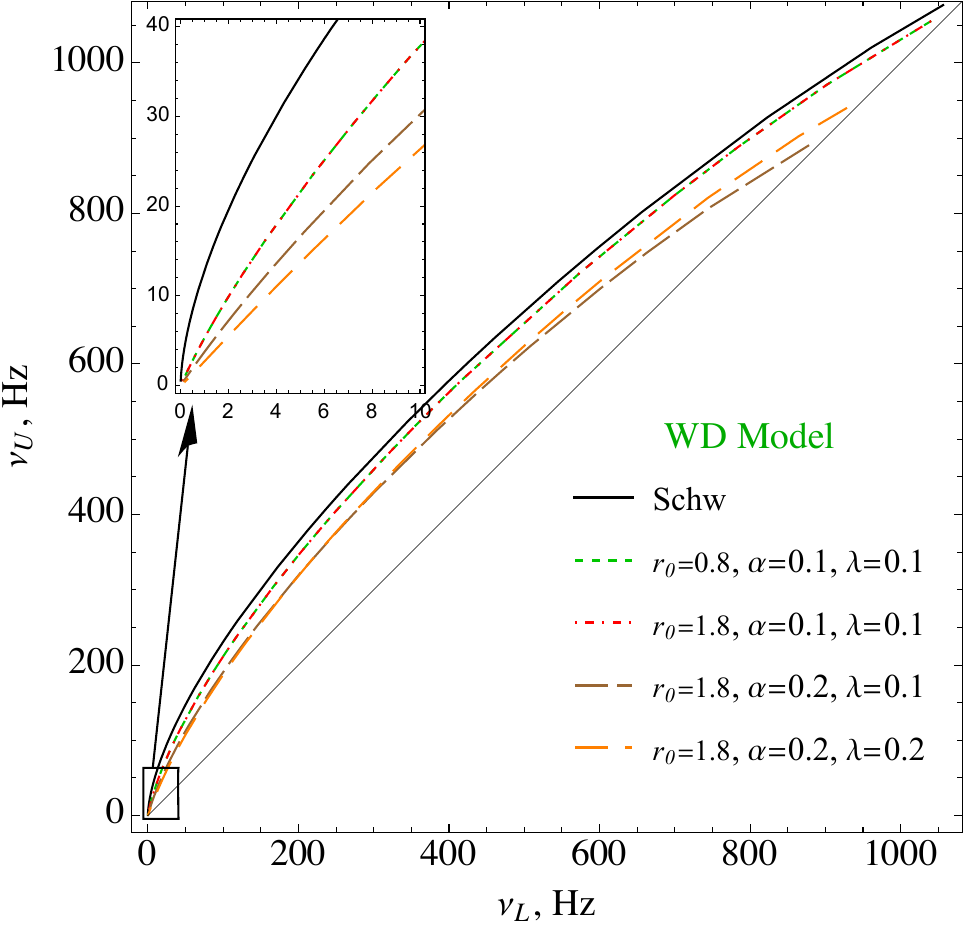}
    \caption{The upper and lower frequency diagram in RP (left panel) and WD (right panel) models.}
    \label{modes}
\end{figure*}

Figure~\ref{modes} presents the relation between the upper and lower QPO frequencies, $\nu_U$ and $\nu_L$, in the RP and WD models for different values of the spacetime deformation parameter $\alpha$, the PFDM parameter $\lambda$, and the scale $r_0$. 
The Schwarzschild configuration is included as a reference, enabling a direct identification of deviations produced by environmental and geometric modifications. In both panels, the modified curves lie below the Schwarzschild prediction. This indicates that the presence of PFDM and spacetime deformation reduces the characteristic orbital and epicyclic frequencies of particles. Such a reduction is associated with an outward displacement of the stability region, where resonance conditions are achieved at lower frequencies. Within the RP model, the deviation becomes stronger as $\alpha$ and $\lambda$ increase. 
The deformation parameter governs the main global shift of the curves, while the PFDM term adjusts the magnitude of the effect. The sensitivity is particularly evident in the low–frequency domain shown in the inset, where even small parameter variations noticeably change the $\nu_U$–$\nu_L$ correlation.
In the WD model, the environmental contribution plays a more competitive role. An increase in $\lambda$ leads to a further decrease in both frequencies, reflecting a modification of the balance between radial and vertical oscillations in the disk. As a result, the resonance conditions responsible for the warped–disk mechanism are significantly altered.

\section{MCMC analyses}\label{sec:7}

Modern X-ray observatories employ highly sensitive detectors capable of measuring the arrival times of X-ray photons with extraordinary temporal resolution, ranging from microseconds to nanoseconds \cite{Gendreau2016NICER}. Quasi-periodic oscillations manifest as distinct, periodic features in the X-ray power density spectra of accreting sources. These spectra are typically obtained by applying Fourier transformation techniques to the photon count rate, allowing the identification of characteristic oscillation frequencies \cite{vanderKlis1989}.  

In this study, we select the test black holes candidates located in microquasars, specifically: XTE J1550-564, GRO J1655--40, and GRS 1915+105 \cite{Orosz2011,Strohmayer2001,Miller2015}. In addition, we include M82 X-1, an ultraluminous X-ray source in the galaxy M82, which is considered a strong candidate for hosting an intermediate-mass black hole \cite{Pasham2014}. The detection of high-frequency QPOs (HF-QPOs) in these systems provides critical evidence supporting the presence of intermediate-mass black holes and offers an opportunity to probe the accretion physics in extreme gravitational environments \cite{Pasham2014,Stuchlik2015M82,Casella2008,Dewangan2006}.  

Studying the frequencies of these QPOs, along with their characteristic ratios, offers valuable insight into the spacetime geometry, mass, and spin of the black holes, as well as the underlying physics of the accretion disk. In this section, we exploit the observational data from the aforementioned sources to derive constraints on the black hole mass and relevant metric parameters. A summary of the sources and the observed pairs of QPO frequencies used in this analysis is provided in Table~\ref{table1}.


\begin{table*}[ht!]\centering
\begin{center}
\caption[]{\label{table1}Frequencies of twin peak QPOs in microquasars and Galactic centre}
\renewcommand{\arraystretch}{1.2}
\begin{tabular}{| l || c  c | c  c | l |}
\hline
Source 
&  $\nu_{\rm{U}}\,$[Hz]&$\Delta\nu_{\mathrm{U}}\,$[Hz]& $\nu_{\rm {L}}\,$[Hz]&$\Delta\nu_{\rm{L}}\,$[Hz]& Mass [\,M$_{\odot}$\,] \\
\hline
\hline
XTE~J1550-564 \cite{Orosz2011} & 276&$\pm\,3$& 184&$\pm\,5$&  ${9.1\pm 0.61}$\\
GRO~J1655--40 \cite{Strohmayer2001}    & 451&$\pm\,5$& 298&$\pm\,4$ & {5.4$\pm$0.3}   \\
GRS~1915+105 \cite{Miller2015}   & 168&$\pm\,3$& 113&$\pm\,5$&  ${12.4^{+2.0}_{-1.8}}$   \\
M82 X-1 \cite{Pasham2014}  & 5.07 & $\pm\,0.06$ & 3.32& $\pm\,0.06$ &$415\pm\,63$\\

\hline
        \end{tabular}
\end{center}
\end{table*}

To obtain the estimation for the five parameters as the peak frequencies of QPOs observed in the microquasars, we perform the $\chi^2$ analysis with \cite{Bambi2015GRO}
\begin{eqnarray}
\chi^{2}(M,r_0,\alpha,\lambda,r)&=&\frac{(\nu_{1\phi}-\nu_{1\rm U})^{2}%
}{\sigma_{1\rm U}^2}+\frac{(\nu_{1\rm per}-\nu_{1\rm L})^{2}%
}{\sigma_{1 \rm L}^2} 
+\frac{(\nu_{1\rm nod}-\nu_{1\rm C})^{2}%
}{\sigma_{1\rm C}^2}+\frac{(\nu_{2\phi}-\nu_{2\rm U})^{2}}%
{\sigma_{2\rm U}^2}
+\frac{(\nu_{2\rm nod}-\nu_{2\rm C})^{2}}%
{\sigma_{2\rm C}^2}~.
\end{eqnarray}

The best-estimated values of the parameters $M, \,r_0,\,\alpha,\,\lambda$ and $r$ in which $\chi_{\rm min}^{2}$ take minimum. The range of the parameters at the confidence level (C.L.) can be determined in the interval $\chi_{\rm min}^{2}\pm \Delta\chi^{2}$. To find these best-estimated values, we select more than 100 values for each parameter, which for these values satisfy the condition $\chi^2\le1$ . After that, we can obtain the mean value ($\mu$) and the standard deviation value ($\sigma$) for $M, \,r_0,\,\alpha,\,\lambda$, $r$ parameters. Obtained values are given in Tab. \ref{prior}.

\begin{table*}[ht!]\centering
\begin{center}
\renewcommand\arraystretch{1.5} 
\caption{\label{prior}%
The Gaussian priors for the black hole mass, $r_0,\,\alpha,\,\lambda$, and the radius of the QPO orbit for the selected objects.}
\begin{tabular}{lcccccccccc}
\hline\hline
\multirow{2}{*}{\textbf{RP}} & \multicolumn{2}{c}{XTE J1550-564} & \multicolumn{2}{c}{GRO J1655-40} & \multicolumn{2}{c}{GRS 1915+105} & \multicolumn{2}{c}{M82 X-1} \\
& $\mu$ & \multicolumn{1}{c}{$\sigma$} & $\mu$          & $\sigma$ & $\mu$ & \multicolumn{1}{c}{$\sigma$} & $\mu$ &\\
\hline
     $M/M_{\odot}$ & $5.601$ & $0.992$ & $3.747$ & $0.466$ & $8.032$ & $0.563$ & $288.69$ & $59.55$ &\\
    
     $r_0$ & $1.051$ & $0.466$   & $1.076$  & $0.427$ & $0.984$ & $0.478$ & $1.036$ & $0.388$  \\ 
    
     $\alpha$ & $0.128$ & $0.078$ & $0.095$ & $0.034$ & $0.178$ & $0.075$ & $0.144$ & $0.047$ \\
     
    $\lambda$ & $0.311$ & $0.162$ & $0.287$ & $0.143$ & $0.351$ & $0.165$ & $0.306$ & $0.095$ \\
    
     $r$ & $6.810$ & $0.654$ & $6.436$ & $0.479$ & $7.266$ & $0.361$ & $7.099$ & $0.843$ \\
     \hline\hline

\multirow{2}{*}{\textbf{WD}} & \multicolumn{2}{c}{XTE J1550-564} & \multicolumn{2}{c}{GRO J1655-40} & \multicolumn{2}{c}{GRS 1915+105} & \multicolumn{2}{c}{M82 X-1} \\
& $\mu$ & \multicolumn{1}{c}{$\sigma$} & $\mu$          & $\sigma$ & $\mu$ & \multicolumn{1}{c}{$\sigma$} & $\mu$ &\\
\hline
     $M/M_{\odot}$ & $7.445$ & $0.732$ & $7.438$ & $0.448$ & $9.904$ & $1.602$ & $353.02$ & $48.01$ &\\
    
     $r_0$ & $1.037$ & $0.484$   & $0.895$  & $0.362$ & $1.004$ & $0.481$ & $0.990$ & $0.375$  \\ 
    
     $\alpha$ & $0.055$ & $0.035$ & $0.082$ & $0.028$ & $0.129$ & $0.071$ & $0.085$ & $0.048$ \\
     
    $\lambda$ & $0.246$ & $0.109$ & $0.227$ & $0.115$ & $0.271$ & $0.142$ & $0.244$ & $0.098$ \\
    
     $r$ & $7.392$ & $0.355$ & $7.602$ & $0.385$ & $8.487$ & $0.796$ & $8.065$ & $0.553$ \\     \hline\hline

\end{tabular}
\end{center}
\end{table*}

We use the Python library \texttt{emcee} \cite{emcee} to perform the MCMC analysis and constrain the parameters of a charged particle around above mentioned black holes. Our analysis are performed in the RP and WD QPO models (see Section \ref{sec:5}).

The posterior distribution, as usual, can be defined as follows \cite{Liu-etal2023},
\begin{eqnarray}
\mathcal{P}(\theta |\mathcal{D},\mathcal{M})=\frac{P(\mathcal{D}|\theta,\mathcal{M})\ \pi (\theta|\mathcal{M})}{P(\mathcal{D}|\mathcal{M})},
\end{eqnarray}
where $\pi(\theta)$ is the prior and $P(D|\theta,M)$ is the likelihood. We choose our priors to be (normal) Gaussian distributions within suitable boundaries (see Table \ref{prior}), \textit{i.e.,} 
$\pi(\theta_i) \sim \exp\left[{\frac{1}{2}\left(\frac{\theta_i - \theta_{0,i}}{\sigma_i}\right)^2}\right]$, \ $\theta_{\text{low},i}<\theta_i<\theta_{\text{high},i}$. In this work, the parameters are $\theta_i=\{M,\,r_0,\,\alpha,\,\lambda,\,r\}$ and $\sigma_i$ are their corresponding variances. Eventually, the likelihood function $\lambda$ can be expressed as
\begin{eqnarray}
\log \lambda = \log \lambda_{\rm U} + \log \lambda_{\rm L},\label{likelyhood}
\end{eqnarray}
where $\log \lambda_{\rm U}$ denotes the likelihood of the astrometric upper frequencies,
\begin{eqnarray}
 \log \lambda_{\rm U} &=& - \frac{1}{2} \sum_{i} \bigg(\frac{\nu_{U\rm, obs}^i -\nu_{U\rm, th}^i}{\sigma^i_{U,{\rm obs}}} \bigg)^2\ ,
\end{eqnarray}
and $\log \lambda_{\rm L}$ represents the likelihood of the data of the lower frequency.

\begin{eqnarray}
 \log \lambda_{\rm L} &=& - \frac{1}{2} \sum_{i} \bigg(\frac{\nu_{L\rm, obs}^i -\nu_{L\rm, th}^i}{\sigma^i_{L,{\rm obs}}} \bigg)^2\ . 
\end{eqnarray}
Here $\nu^i_{U,\rm obs}$, $\nu^i_{L,\rm obs}$ are direct observational results of the upper and lower frequencies of selected objects and $\sigma^i_{U,{\rm obs}}, \quad\sigma^i_{L,{\rm obs}}$ represent the uncertainty (error) in the measured frequencies. On the other side, $\nu^i_{U,\rm th}$, $\nu^i_{L,\rm th}$ are the respective theoretical estimations.

Next, we perform the MCMC simulation to constrain the parameters ($M,\,r_0,\,\alpha,\,\lambda,\,r$) for the Dymnikova black hole immersed in PFDM. We sampled approximately $5\times 10^4$ points for each parameter from a Gaussian prior distribution, allowing us to explore the physically possible parameter space within set boundaries and identify the best-fitting parameter values.

\begin{figure}[tbhp]
    \centering
    \includegraphics[width=0.43\linewidth]{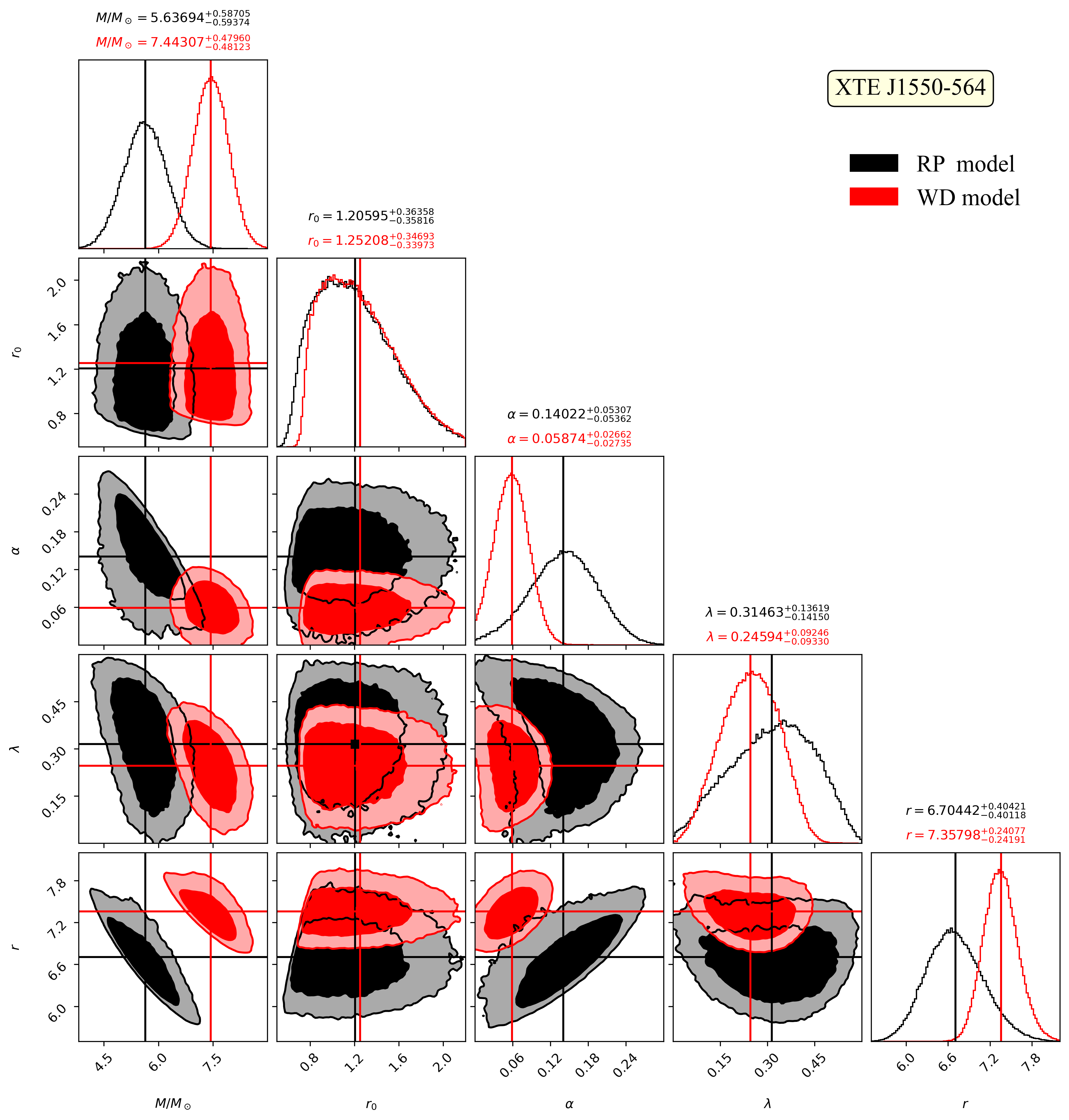}
    \includegraphics[width=0.43\linewidth]{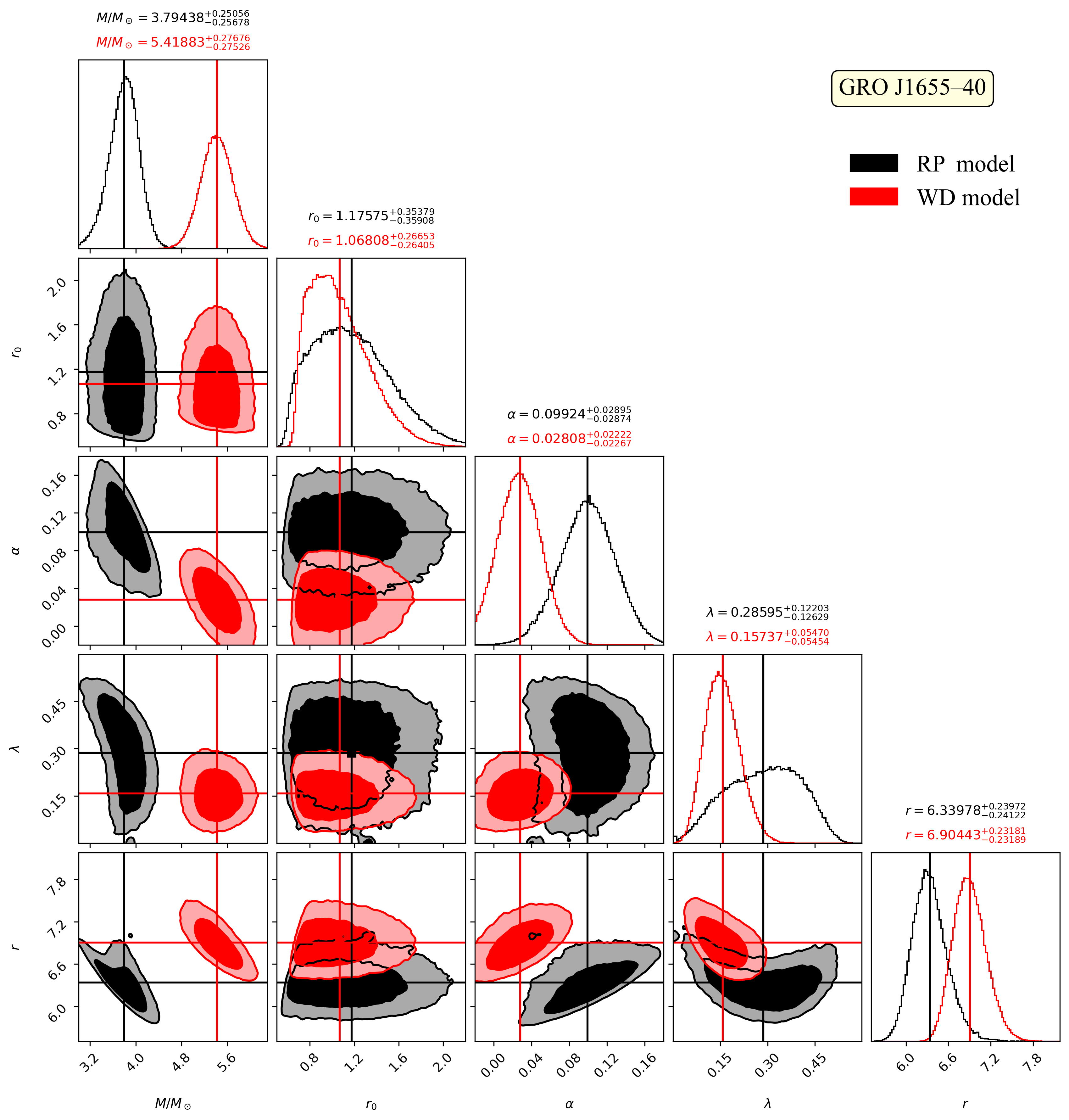}
    \includegraphics[width=0.43\linewidth]{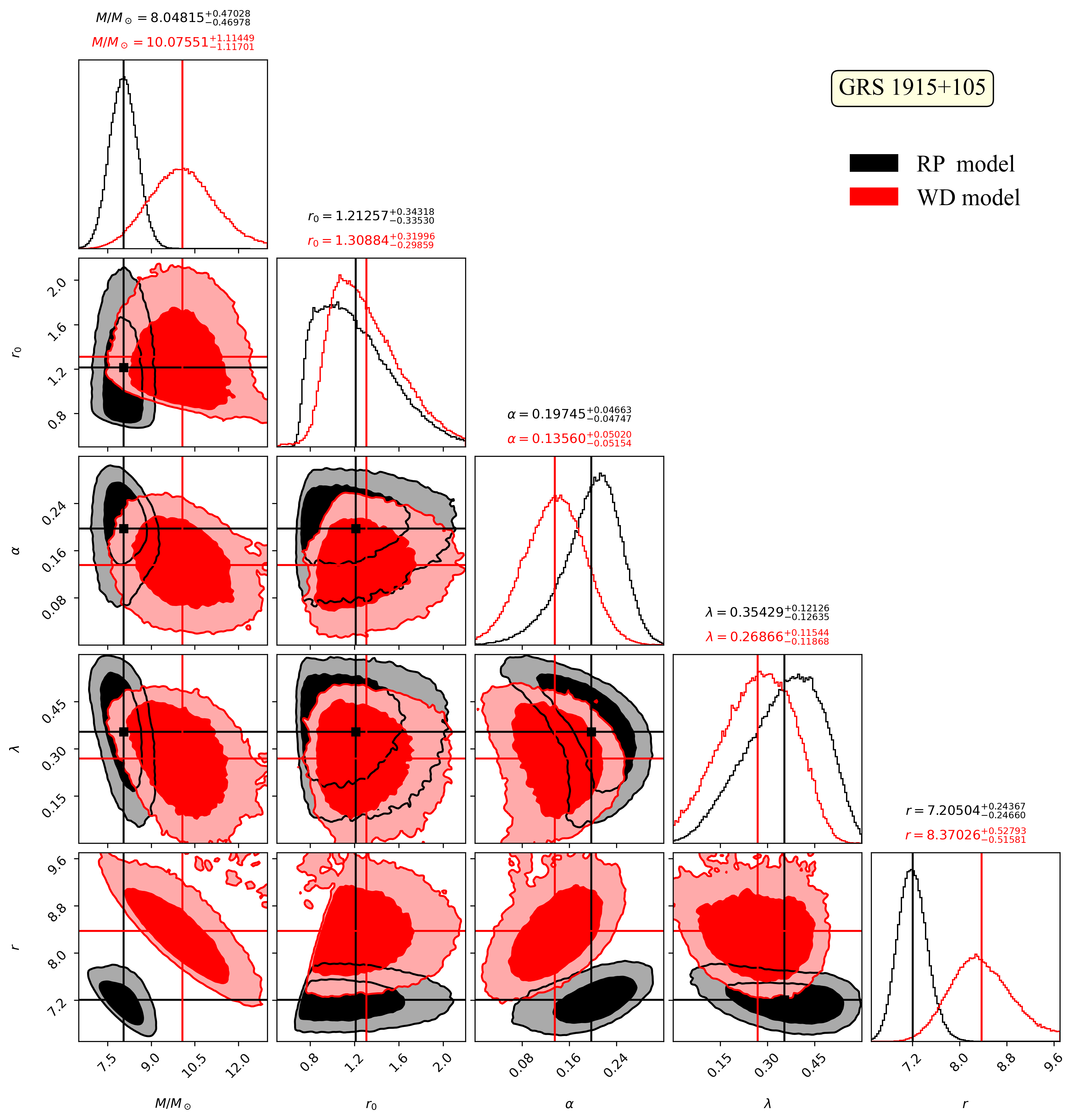}
    \includegraphics[width=0.43\linewidth]{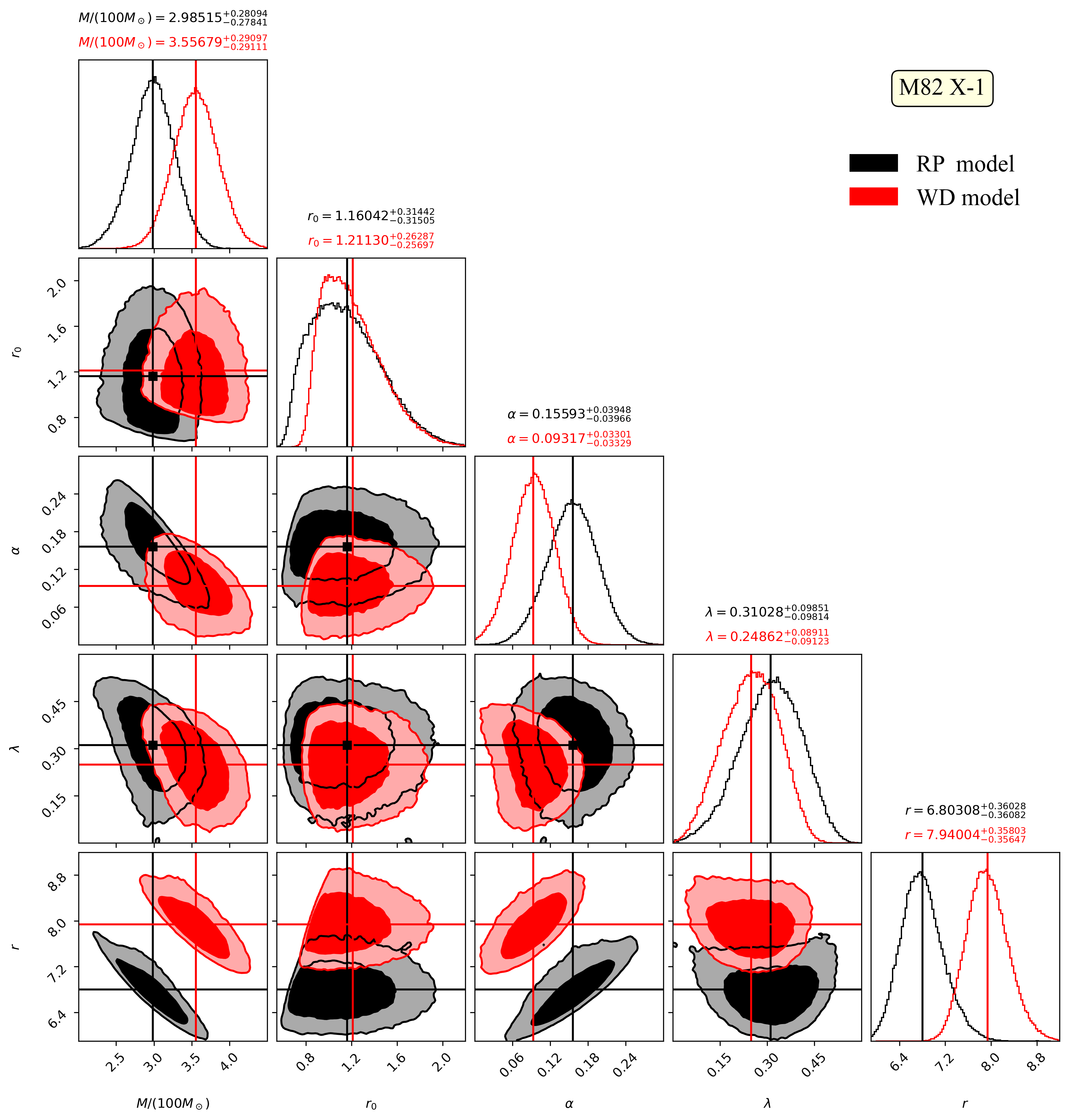}
    \caption{Constraints on the black hole mass, $r_0,\,\alpha,\,\lambda$, and the radius of the QPO orbit from a four-dimensional MCMC analysis $1\sigma$ and $2\sigma$ using the QPO data for the stellar-mass black hole XTE J1550-564 (top left), GRO J1655–40 (top right), GRS 1915+105 (bottom left) and M82 X-1 (bottom right) in the selected models. \label{XTE}}
    \label{fig:1}
\end{figure}

Now, we collect the best-fit values for the parameters $M,\,r_0,\,\alpha,\,\lambda,\,r$ obtained by MCMC results in Figs.\ref{XTE}  and present them all below in a table form.

\begin{table*}\centering
\centering
\renewcommand{\arraystretch}{1.5}
\setlength{\tabcolsep}{8pt}
\begin{tabular}{|l|p{3cm}|p{2cm}|p{2cm}|p{2cm}|p{2cm}|}
\hline
\textbf{Model: RP} & ${M/(M_\odot)}$ & $r_0$ & $\alpha$ & $\lambda$ & $r/M$ \\
\hline
XTE~J1550-564 & $5.636^{+0.587}_{-0.593}$ & \(1.206^{+0.363}_{-0.358}\)  & \(0.140^{+0.053}_{-0.053}\)  & \(0.316^{+0.136}_{-0.141}\) &\(6.704^{+0.404}_{-0.401}\) \\

\hline
GRO~J1655--40 & \(3.795^{+0.250}_{-0.256}\) & \(1.176^{+0.353}_{-0.359}\) & \(0.099^{+0.028}_{-0.028}\) & \(0.286^{+0.122}_{-0.126}\) &\(6.339^{+0.239}_{-0.241}\)\\

\hline
GRS~1915+105 & \(8.048^{+0.470}_{-0.469}\) & \(1.212^{+0.343}_{-0.335}\) & \(0.197^{+0.046}_{-0.047}\) & \(0.354^{+0.121}_{-0.126}\) &\(7.205^{+0.243}_{-0.246}\)\\

\hline
M82 X-1 & \( 298.51^{+28.09}_{-27.81}\) & \(1.160^{+0.314}_{-0.315}\) & \(0.155^{+0.039}_{-0.039}\) & \(0.310^{+0.098}_{-0.098}\) &\(6.803^{+0.360}_{-0.360}\)\\

\hline
\textbf{Model: WD} &  &  &  & & \\
\hline
XTE~J1550-564 & \(7.443^{+0.479}_{-0.481}\) & \(1.206^{+0.346}_{-0.339}\)  & \(0.058^{+0.026}_{-0.027}\)  & \(0.246^{+0.092}_{-0.093}\) &\(7.358^{+0.241}_{-0.242}\)\\

\hline
GRO~J1655--40 & \(5.419^{+0.276}_{-0.275}\) & \(1.175^{+0.353}_{-0.359}\) & \(0.028^{+0.022}_{-0.022}\) & \(0.157^{+0.054}_{-0.054}\) &\(6.904^{+0.232}_{-0.232}\) \\
\hline
GRS~1915+105 & \(10.075^{+1.114}_{-1.117}\) & \(1.308^{+0.319}_{-0.298}\) & \(0.135^{+0.050}_{-0.051}\) & \(0.268^{+0.115}_{-0.118}\) &\(8.370^{+0.527}_{-0.515}\)\\
\hline
M82 X-1 & \(355.68^{+29.09}_{-29.11}\) & \(1.211^{+0.263}_{-0.256}\) & \(0.093^{+0.033}_{-0.033}\) & \(0.248^{+0.089}_{-0.091}\) &\(7.940^{+0.358}_{-0.356}\)\\
\hline
\end{tabular}
\caption{Best-fit parameters from various sources for different QPO models.}
\label{tab3}
\end{table*}

The best-fit parameters in Table~\ref{tab3} show different black hole masses across the RP and WD models. We can conclude the following from Tab.\ref{tab3} and Fig. \ref{XTE} :
\begin{itemize}
\item The MCMC analysis reveals a systematic discrepancy between the inferred best--fit mass values obtained within the RP and WD models. For all four selected black hole candidates, the best--fit mass derived in the RP model is consistently smaller, whereas the WD model yields comparatively larger mass estimates. This uniform trend across all sources indicates that the observed mass shift does not originate from the intrinsic physical nature of the black holes themselves. Instead, it arises from structural differences between the underlying test models (RP and WD). 
\item The posterior distributions of the parameters $M$, $\alpha$, $\lambda$, and $r$ exhibit shapes that are well approximated by Gaussian profiles, indicating stable convergence and weak non-linear degeneracies within the explored parameter space. In contrast, the posterior behavior of the parameter $r_{0}$ significantly deviates from Gaussian distribution. For all considered black holes, the $1\sigma$ interval of $r_{0}$ is consistently confined within the range
$r_{0} \in (0.4,\, 1.7)$ This systematic clustering implies that the parameter region $r_{0} \in (0,\, 0.4)$ is strongly not acceptable by the MCMC analysis. Consequently, selecting $r_{0}$ within this lower interval appears to be incompatible with the QPO observational constraints.
\item The posterior analysis of the radial coordinate of test particles yields the following best--fit intervals. For RP model, the preferred region is $r_{\text{best}} \in (6.34,\, 7.21)$ whereas in the WD model the corresponding interval is  $r_{\text{best}} \in (6.90,\, 8.37).$ For all investigated sources and model realizations, it is evident that $r_{\mathrm{ISCO},\,\text{best}} < r_{\text{best}}$ as illustrated in Fig.~3. This  indicates that the preferred orbital radii associated with the QPO fits are located outside the innermost stable circular orbit, ensuring dynamical stability of the test particle motion within both model frameworks.
\end{itemize}

\section{Conclusion}\label{sec:8}

In this work, we have investigated in detail the physical properties of a Dymnikova black hole immersed in a perfect fluid dark matter background, characterized by the parameter $\lambda$, and additionally surrounded by a cloud of strings described by the parameter $\alpha$. The combined influence of these two external fields modifies both the geometrical structure and the dynamical behavior of the spacetime. In particular, we analyzed the thermodynamic properties of the system—including the Hawking temperature and the specific heat capacity-as well as the dynamics of massless and massive test particles, quasi-periodic oscillations, and the resulting black hole shadow profiles.

From the thermodynamic perspective, the presence of perfect fluid dark matter and string-like sources produces significant deviations from the standard Dymnikova solution. Both the Hawking temperature and the heat capacity depend explicitly on the parameters $\lambda$ and $\alpha$, while the standard Dymnikova black hole is naturally recovered in the limit $\lambda = 0 = \alpha$. The Hawking temperature exhibits a non-monotonic behavior: it increases with the event horizon radius, reaches a maximum at a critical radius, and then decreases for larger radii. The peak value of the temperature grows as either $\alpha$ or $\lambda$ increases, indicating that the surrounding matter fields enhance the thermal activity of the black hole. The heat capacity further reveals the existence of parameter-dependent phase transitions, signaling changes in local thermodynamic stability.

The analysis of photon motion shows that PFDM and the cloud of strings substantially modify the photon sphere structure and, consequently, the size of the black hole shadow. For a fixed $\alpha$, increasing $\lambda$ reduces both the photon sphere radius and the shadow radius. In contrast, for a fixed negative $\lambda$, increasing $\alpha$ enlarges these radii. This demonstrates that the two parameters play competing roles in determining the optical appearance of the black hole.

For massive particle motion, both the specific energy and the specific angular momentum of circular orbits decrease as $\alpha$ or $\lambda$ increases. This reflects a modification of the effective gravitational potential and alters the stability conditions of particle motion, with potential implications for accretion disk dynamics and high-energy astrophysical phenomena.

In addition, we investigated the QPO properties of the Dymnikova black hole immersed in PFDM within two widely used theoretical frameworks: the relativistic precession and warped disk models. The presence of spacetime deformation ($\alpha$) and environmental effects ($\lambda$) systematically reduces the characteristic orbital and epicyclic frequencies compared to the Schwarzschild case. This reduction corresponds to a modification of the effective potential and an outward displacement of the resonance region responsible for twin-peak QPO generation.

Using observational data from XTE J1550-564, GRO J1655–40, GRS 1915+105, and M82 X-1, we performed detailed $\chi^2$ and MCMC analyses to constrain the black hole mass and metric parameters. The posterior distributions show stable convergence for $M$, $\alpha$, $\lambda$, and the orbital radius $r$, while the parameter $r_0$ is restricted to a viable range, with values below $r_0 \lesssim 0.4$ strongly disfavored by QPO observations. A systematic discrepancy in the inferred mass values is observed between the RP and WD models: the RP model consistently yields lower mass estimates, whereas the WD model predicts higher values. This difference originates from structural distinctions in the frequency identifications of the two models rather than from intrinsic physical differences of the black holes.

The preferred orbital radii obtained from the fits lie outside the innermost stable circular orbit, ensuring dynamical stability and consistency with accretion disk physics. The inferred parameter ranges remain physically reasonable for both stellar-mass and intermediate-mass black hole candidates, including M82 X-1.

Overall, the results demonstrate that the combined influence of perfect fluid dark matter and string clouds leads to non-trivial modifications in thermodynamics, particle dynamics, optical properties, and QPO phenomenology of the Dymnikova black hole. QPO timing data provide meaningful constraints on spacetime deformation and dark matter environment parameters, while preserving asymptotic consistency with general relativity. These findings highlight the potential of QPO observations as a sensitive probe of strong-field gravity and environmental effects around compact objects.

\footnotesize

\section*{Acknowledgments}

F.A. acknowledges the Inter University Centre for Astronomy and Astrophysics (IUCAA), Pune, India for granting visiting associateship.

\section*{Data Availability Statement}

No new data were generated or analyzed in this manuscript [Authors comments: There is no new data generated in this manuscript].

\section*{Code/Software Statement}

No code or software were generated in this manuscript [Authors comments: There is no new code or software in this manuscript].

\end{document}